\documentclass[notitlepage,floats,aps,nofootinbib,preprintnumbers,superscriptaddress,prd,10pt,balancelastpage,twocolumn]{revtex4-1}
\usepackage{amsmath,amssymb,amsfonts}
\usepackage{hyperref}

\hypersetup{colorlinks,citecolor=blue}
\usepackage{graphicx}
\usepackage{xcolor}
\usepackage{subfigure}
\usepackage{nicefrac}
\usepackage{mathrsfs}
\usepackage{mdframed}
\usepackage{ulem}
\def\beq{\begin{equation}\begin{aligned}}
\def\eeq{\end{aligned}\end{equation}}

\let\OLDthebibliography\thebibliography
\renewcommand\thebibliography[1]{\OLDthebibliography{#1}
  \setlength{\itemsep}{3pt}}

\newcommand{\lsim}{{\;\raise0.3ex\hbox{$<$\kern-0.75em\raise-1.1ex\hbox{$\sim$}}\;}}
\newcommand{\gsim}{{\;\raise0.3ex\hbox{$>$\kern-0.75em\raise-1.1ex\hbox{$\sim$}}\;}}

\newcommand{\mb}[1]{\boldsymbol{#1}}

\def\bea{\begin{equation}\begin{aligned}}
\def\eea{\end{aligned}\end{equation}}
\def\beq{\begin{equation}}
\def\eeq{\end{equation}}
\def\beqnn#1\eeq{\begin{align*}#1\end{align*}}
\def\ba{\begin{array}}
\def\ea{\end{array}}
\def\bc{\begin{center}}
\def\ec{\end{center}}

\def\Trh{T_{\rm rh}}
\def\Mpl{M_{\rm Pl}}

\newcommand{\GeV}{\mathrm{\;GeV}}
\newcommand{\TeV}{\mathrm{\;TeV}}

\newcommand{\mpl}{M_{\rm Pl}}

\newcommand{\RUV}{R_\textrm{UV}}
\newcommand{\cUV}{c_\textrm{UV}}

\newcommand{\Fig}[1]{Figure~\ref{#1}}

\newcommand{\Eq}[1]{eq.~(\ref{#1})}

\begin{document}

\title{Parametric Coincidence in the Baryon to Dark Matter Ratio from\\ Affleck-Dine Baryogenesis and UV Freeze-in Dark Matter}


\author{Jae Hyeok Chang} 
\affiliation{Theory Division, Fermi National Accelerator Laboratory, Batavia, IL 60510, USA}
\affiliation{Department of Physics, University of Illinois Chicago, Chicago, IL 60607, USA}
\author{Chang Sub Shin} 
\affiliation{Department of Physics and Institute of Quantum Systems, Chungnam National University, Daejeon 34134, South Korea}
\affiliation{Center for Theoretical Physics of the Universe, Institute for Basic Science, Daejeon 34126, South Korea}
\affiliation{Korea Institute for Advanced Study, Seoul 02455, South Korea}
\author{James Unwin} 
\affiliation{Department of Physics, University of Illinois Chicago, Chicago, IL 60607, USA}
\preprint{FERMILAB-PUB-25-0376-T-V}

\begin{abstract} 
We highlight that the observed concurrence between the baryon and dark matter relic densities can be explained via a parametric coincidence between two distinct production mechanisms: Affleck-Dine baryogenesis and dark matter UV freeze-in. In the Affleck-Dine mechanism, the baryon asymmetry is naturally proportional to the inflationary reheating temperature $\Trh$, which also plays a critical role in setting the relic abundance of UV freeze-in dark matter. Since Affleck-Dine baryogenesis requires flat directions in the potential, the framework is inherently supersymmetric, offering compelling UV freeze-in dark matter candidates such as the gravitino. We outline scenarios in which $\Trh$ simultaneously determines both relic abundances, resulting in a baryon-to-dark matter ratio of order unity that is largely insensitive to $\Trh$. 
We also discuss the conditions required to avoid Q-ball formation or dark matter production by other mechanisms, such as NLSP decays, to preserve the parametric coincidence between baryon and dark matter abundances.

\end{abstract}

\maketitle

\section{introduction}

The genesis of matter remains a fundamental question, in particular, the dual origins of dark matter and the matter-antimatter asymmetry. This includes the curious similarity between dark matter and baryon densities: $\Omega_{\rm DM}\simeq 5 \Omega_{\rm B}$ \cite{Aghanim:2020}. For scenarios in which dark matter and the baryon asymmetry are produced at or before the onset of the radiation-dominated era, cosmological parameters such as the reheating temperature $T_{\rm rh}$ significantly influence the late-time abundances. 
Different UV-sensitive production mechanisms for dark matter and the baryon asymmetry 
usually imply that the relative ratio between them would depend on both particle physics model parameters and the cosmological parameters. Upon closer examination of this reheating temperature dependence, intriguing coincidences arise. 

 UV freeze-in   \cite{Hall:2009bx,Elahi:2014fsa} relies on out-of-equilibrium production of dark matter from a thermal bath, with production rates increasing at higher temperatures. 
The production peaks at the beginning of the radiation-dominated era following inflationary reheating. This occurs through higher-dimensional operators. Specifically, when dimension-five operators are involved, the dark matter abundance scales linearly with reheating temperature $\Omega_{\rm DM}\propto T_{\rm rh}$. 
On the other hand, in the Affleck-Dine mechanism \cite{Affleck:1985} where baryon asymmetry is generated through a rotating scalar field during reheating the baryon-to-photon ratio also exhibits a linear dependence on the reheating temperature, 
$\Omega_{\rm B}\propto T_{\rm rh}$. This is
 due to the dilution effect caused by entropy production prior to the end of reheating.
Despite the distinct production processes, this linear dependence on $T_{\rm rh}$ cancels out in their relative abundance, allowing their ratio $\Omega_{\rm DM}/\Omega_\textrm{B}$ to remain independent of $T_{\rm rh}$, and resulting in models that have predictive features despite their UV sensitivity.

Notably, both the Affleck-Dine mechanism and UV freeze-in dark matter find a natural framework within supersymmetry (SUSY). The Affleck-Dine mechanism relies on renormalizable flat directions  \cite{Dine:1996}, which are highly fine-tuned in non-supersymmetric settings. 
Moreover, within SUSY extensions of the Standard Model, there are a number of classic dark matter candidates with abundances that are set via UV freeze-in, most prominently the gravitino.

Here we study the prospect of linking the dark matter abundance set by UV freeze-in and the baryon asymmetry generated by the Affleck-Dine mechanism through their shared $T_{\rm rh}$ dependence. It should be noted that the link between  UV freeze-in and Affleck-Dine will be broken if there are other sources of baryon asymmetry or dark matter production. For instance, sparticle decays to the lightest SUSY particle (LSP) after decoupling, or if Q-ball formation is effective, these can disconnect the dark matter and/or the baryon asymmetry from $T_{\rm rh}$. Here we identify models for which these issues are avoided.

This paper is structured as follows in Section~\ref{Sec2} we recall the process of Affleck-Dine baryogenesis and the UV freeze-in mechanism, highlighting the general conditions under which one can simultaneously match the observed $\Omega_{\rm DM}$ and $\Omega_{B}$ via an appropriate choice of $T_{\rm rh}$. Section~\ref{Sec3} discusses UV freeze-in within the context of supersymmetry. Section \ref{Sec4} identifies Affleck-Dine scenarios that avoid Q-balls, which otherwise disrupt the coincidence between $\Omega_{\rm DM}$ and $\Omega_{B}$. Finally, in Section~\ref{Sec6} we provide two specific examples that realize our ambition of $T_{\rm rh}$ insensitivity in $\Omega_{\rm DM}/\Omega_{B}$, in which the Affleck-Dine field is identified as either a RH sneutrino or an R-parity violating operator.  Concluding remarks are presented in Section~\ref{Sec7}.

\section{parametric dependences of Dark Matter and the baryon asymmetry}
\label{Sec2}

We are interested in settings in which the late-time abundances of dark matter and baryons have linear dependences on the reheating temperature of the Universe. Before discussing specific models, in this section, we will first explore the general features of both UV freeze-in dark matter and Affleck-Dine baryogenesis, and discuss how these models realize our scenario.

\subsection{UV freeze-in DM production}\label{sec2:UVFI}

Let us consider a fermion \(\chi\), representing dark matter, which is a singlet under both the gauge group of the Standard Model and baryon and lepton numbers. The stability of dark matter is guaranteed by a \(Z_2\) symmetry under which \(\chi \to -\chi\). In this setup, the leading operators between the dark matter and the Standard Model particles are mass dimension-five operators, such as
\beq \label{eq:UVFI_Higgs}
\frac{1}{\Lambda} \chi \chi O_{\rm SM}, \quad \frac{1}{\Lambda} \chi X^a O^a_{\rm SM}   .
\eeq
Here, $O_{\rm SM}$ is a Standard Model singlet dimension-two operator involving visible sector fields. The second dimension-five operator can also exist if there exists another heavier $Z_2$ charged fermion \(X^a\) that carries an appropriate Standard Model charge. The cutoff scale \(\Lambda\), is an effective scale influenced by UV physics.

For dimension-five operators with a sufficiently large cutoff scale $\Lambda$, the dark matter abundance can be determined by the UV freeze-in mechanism \cite{Hall:2009bx,Elahi:2014fsa}.
A natural choice for $\Lambda$ is the reduced Planck mass, $\Lambda = \Mpl$.
In this case, dark matter remains out of equilibrium, and its production is dominated by direct thermal processes at high temperatures.
At temperatures well above the masses of the relevant particles, the Boltzmann equation governing the dark matter number density takes the form
\begin{equation} \label{eq:UVFIthp}
\dot{n}_\chi + 3 H n_\chi = \RUV \frac{T^6}{\Mpl^2},
\end{equation}
where $\RUV$ is a model-dependent $\mathcal{O}(1)$ parameter associated with UV freeze-in (which may exhibit some mild temperature-dependence).
It follows that the dark matter yield $Y_\chi$ is 
\bea\label{eq:nchi}
Y_\chi = \frac{n_{\chi}}{s}  =  \cUV \RUV \left(\frac{T_{\rm rh}}{M_{\rm Pl}}\right),
\eea 
where $\cUV$ is a numerical factor given by
\bea \label{eq:cchi}
\cUV &= \frac{1}{\sqrt{2}} \left(\frac{3\sqrt{5}}{\pi}\right)^3 \frac{\tilde{g}(\Trh)}{g_s^*(\Trh) \sqrt{g^*(\Trh)}}\\
&\approx 4.9 \times 10^{-4} \left( \frac{200}{g^*(\Trh)} \right)^{3/2}   ,
\eea
where $g^*$ ($g_S^*$) is the relativistic degree of freedom for the energy (entropy) density, and we introduce
\beq
\tilde{g} \equiv 1+\frac{T}{3}\frac{d \log(g_S^*)}{dT}.
\eeq
 In the second line of \Eq{eq:cchi}, we take $\tilde{g}\approx 1$ and approximate $g_S^*\approx g^*$.
In the case that $\chi$ comprises 100\% of the dark matter abundance, we have the dark matter yield $Y_\textrm{DM}=Y_\chi$.
Observe, in particular, the linear dependence on the reheating temperature $T_{\rm rh}$ in the dark matter yield in \Eq{eq:nchi}.

\subsection{Affleck-Dine mechanism}

The Affleck-Dine mechanism  \cite{Affleck:1985}  for generating a baryon asymmetry relies on the dynamics of a scalar field carrying an approximately conserved quantum number, e.g.~\(B-L\). Let us denote this complex scalar field as \(\phi\), with a corresponding U(1)\({}_\phi\) global transformation under which \(\phi \rightarrow e^{ic} \phi\). This global symmetry is assumed to be slightly broken,  typical at the Planck scale $M_{\rm Pl}$, with the breaking effect increasing as \(|\phi|\) takes on larger values. Decomposing \(\phi\) into fields representing its magnitude and phase gives
\begin{equation}
\phi(x) = \frac{1}{\sqrt{2}} f(x) e^{ i\theta(x)}.
\end{equation}
The associated current from the field dynamics can be expressed as \( (j_\phi)_\mu  = f^2 \partial_\mu\theta \). 

Assuming an initially large expectation value for \(f(x)\) compared to the background temperature, \(\phi\) can be naturally out-of-equilibrium. For a homogeneous field expectation value, the net number density of $\phi$ is
\begin{equation}
\bar{n}_\phi(t) = n_\phi (t) - n_{\phi^*} (t) = f^2(t) \dot\theta(t).
\end{equation}
The Affleck-Dine mechanism describes how, in the early Universe, this large initial \(f\) value and a nonzero {\it kick} velocity \(\dot\theta\) can arise based on the scalar potential 
\beq
V(\phi)= V(|\phi|) + \delta V(f, \theta),
\eeq where $\delta V$ is the global symmetry-breaking contribution. 

The Boltzmann equation governing the evolution of the particle asymmetry $\bar{n}_\phi$ is given by
\begin{equation}\label{eq:ADnumber}
\dot{\bar{n}}_\phi + 3 H \bar{n}_\phi =  -  \frac{\partial}{\partial\theta} \delta V.
\end{equation}
Typically,  one expects \( \partial_\theta   \delta V\sim \epsilon f^n \cos (n\theta)\), where $n \geq 3$ is an integer and $\epsilon$ is a small constant with dimension $4-n$.
At a certain time $t$ in the early Universe, 
the RHS of eq.~(\ref{eq:ADnumber}) becomes effective to kick the $\theta$ field, leading to a non-zero $\bar{n}_\phi$. However, its effect quickly decreases once the scalar field begins a rapid rotation around the origin in the complex field space.  Once the net $\phi$-number is fixed, $\bar{n}_\phi$ scales as $1/a^3$. 
 
In this sense, the Affleck-Dine mechanism resembles UV freeze-in, as the net  \(\phi\)-number is generated and frozen before it reaches the equilibrium value at high temperature, but unlike freeze-in, the Affleck-Dine mechanism is highly non-thermal, which is important for the Sakharov conditions \cite{Sakharov:1967dj}, and it is due to the field nature of the complex scalar field. The net $\phi$-number density generated through the Affleck-Dine mechanism is parametrically 
\bea \label{eq10}
\bar{n}_\phi(t_{\rm AD}) = f^2(t_{\rm AD}) \dot\theta(t_{\rm AD})\quad {\rm at}\  t= t_{\rm AD},
\eea 
where $t_{\rm AD}$ is the time scale at which the potential effectively kicks in and a sizable asymmetry is generated. 
In a typical Affleck-Dine baryogenesis scenario, $t_{\rm AD}$ is set during the early matter-dominated era, i.e, before the end of reheating. This period is driven by a field (e.g., inflaton or curvaton) whose decay into Standard Model particles initiates reheating.  
Thus the net \(\phi\)-number-to-entropy ratio decreases continuously after its generation due to radiation production in the early matter-dominated era. At the temperature \(T = T_{\rm rh}\) (corresponding to time \(t = t_{\rm rh}\)), the early matter density crosses the radiation density (\(\rho_{\rm emat} \sim \rho_{\rm rad}\)), marking the beginning of the radiation-dominated era. 
This assumes that reheating after inflation (due to decays of the inflaton field $I$) is perturbative and completes efficiently, such that $T_{\rm RH}\simeq \sqrt{\Gamma_I M_{\rm Pl}}$ is a faithful approximation.

Prior to this transition, the total energy density scales as \(\rho_{\rm emat} \propto 1/a^3\), similar to \(\bar{n}_\phi\), resulting in
\bea \label{eq:nphi}
\frac{\bar{n}_\phi}{s} & = \left.\frac{\bar{n}_\phi}{\frac{4}{3T}\rho_{\rm rad}}\right|_{T_{\rm rh}} \simeq \left.\frac{3\bar{n}_\phi(t) T_{\rm rh}}{4\rho_{\rm emat}(t)}\right|_{t_{\rm AD}\leq t \leq t_{\rm rh}}\\
&= \left(\frac{f^2(t_{\rm AD}) \dot\theta(t_{\rm AD})}{4 H^2(t_{\rm AD}) M_{\rm Pl}}\right)\left(\frac{T_{\rm rh}}{M_{\rm Pl}}\right).
\eea
Here, the dynamical condition $\rho_{\rm emat}(t_{\rm rh})=\rho_{\rm rad}(t_{\rm rh})$ is used, and, in the lower line, for $t_{\rm AD} \ll t_{\rm rh}$ we substitute   
\begin{equation}
\rho_{\rm emat}(t_{\rm AD})= 3H^2(t_{\rm AD}) M_{\rm Pl}^2~.
 \end{equation}

We note that if a flat direction obtains a very large initial displacement, this can lead to $\bar{n}_\phi\not\propto T_{\rm rh}$, in which case eq.~(\ref{eq:nphi}) will no longer hold and our connection between UV freeze-in and baryogenesis will be spoiled. There are two separate concerns here; the first is that the flat direction may come to dominate the energy density of the Universe, which would lead to an additional entropy injection when it decays. We derive the condition under which matter domination is avoided in Appendix \ref{ApA}. The second issue is that if the VEV is large ($\gtrsim10^{12}$ GeV) and breaks a Standard Model gauge symmetry (e.g.~${\mb L}{\mb H}$ or ${\mb U}{\mb D}{\mb D}$) this will tend to delay thermalization after perturbative inflaton decay \cite{Allahverdi:2005mz}. In addition to lowering the reheat temperature, this can result in pre-thermalization production of dark matter being important \cite{Garcia:2018wtq}. 
 We discussed such effect in Section~\ref{sec:RPVAD}.

Although there is significant freedom in the cosmological parameters $H(t_{\rm AD})$, $\dot\theta(t_{\rm AD})$, and $f(t_{\rm AD})$, these parameters are typically interrelated in specific realizations. For instance, in the context of SUSY, the soft SUSY breaking scale $m_{\rm soft}$ in the visible sector sets the mass scale of $\phi$, while the explicit U(1)${}_{\phi}$ breaking commonly arises due to a contribution in the scalar potential of the form $\frac{m_{\rm soft}}{\mpl} \phi^4$.
This leads to the following relations
\bea\label{sim}
&\dot\theta(t_{\rm AD}) \sim H(t_{\rm AD}) \sim m_{\rm soft},\\ & f^2(t_{\rm AD}) \sim H(t_{\rm AD}) M_{\rm Pl}
\sim m_{\rm soft} M_{\rm Pl}.
\eea 
Using these characteristic values in eq.~(\ref{eq:nphi}) one finds
\beq\label{eq:Yphi}
Y_\phi = \frac{\bar{n}_\phi}{s}  = R_{\rm AD} \left(\frac{T_{\rm rh}}{M_{\rm Pl}}\right)   ,
\eeq
where $R_{\rm AD}$ is a model-dependent coefficient. If one assumes that the quantities in eq.~(\ref{sim}) only differ by an $\mathcal{O}(1)$ number, then the product of these which comprises $R_{\rm AD}$ is such that one reasonably expects this prefactor to fall in the range $0.01\lesssim R_{\rm AD} \lesssim 10$.

Eventually, this net $\phi$-number ($\phi$-$\bar\phi$ asymmetry) is transferred to the asymmetry between the Standard Model quark and anti-quark, leading to the baryon yield today $Y_\textrm{B}\simeq Y_\phi$. 
Moreover, the linear dependence of $Y_\textrm{B}$ on the reheating temperature   $T_{\rm rh}$ can be clearly inferred from eq.~(\ref{eq:Yphi}).

\subsection{Dark matter to baryon ratio}

\begin{figure*}[t!]
\centering
\includegraphics[width=0.85\textwidth]{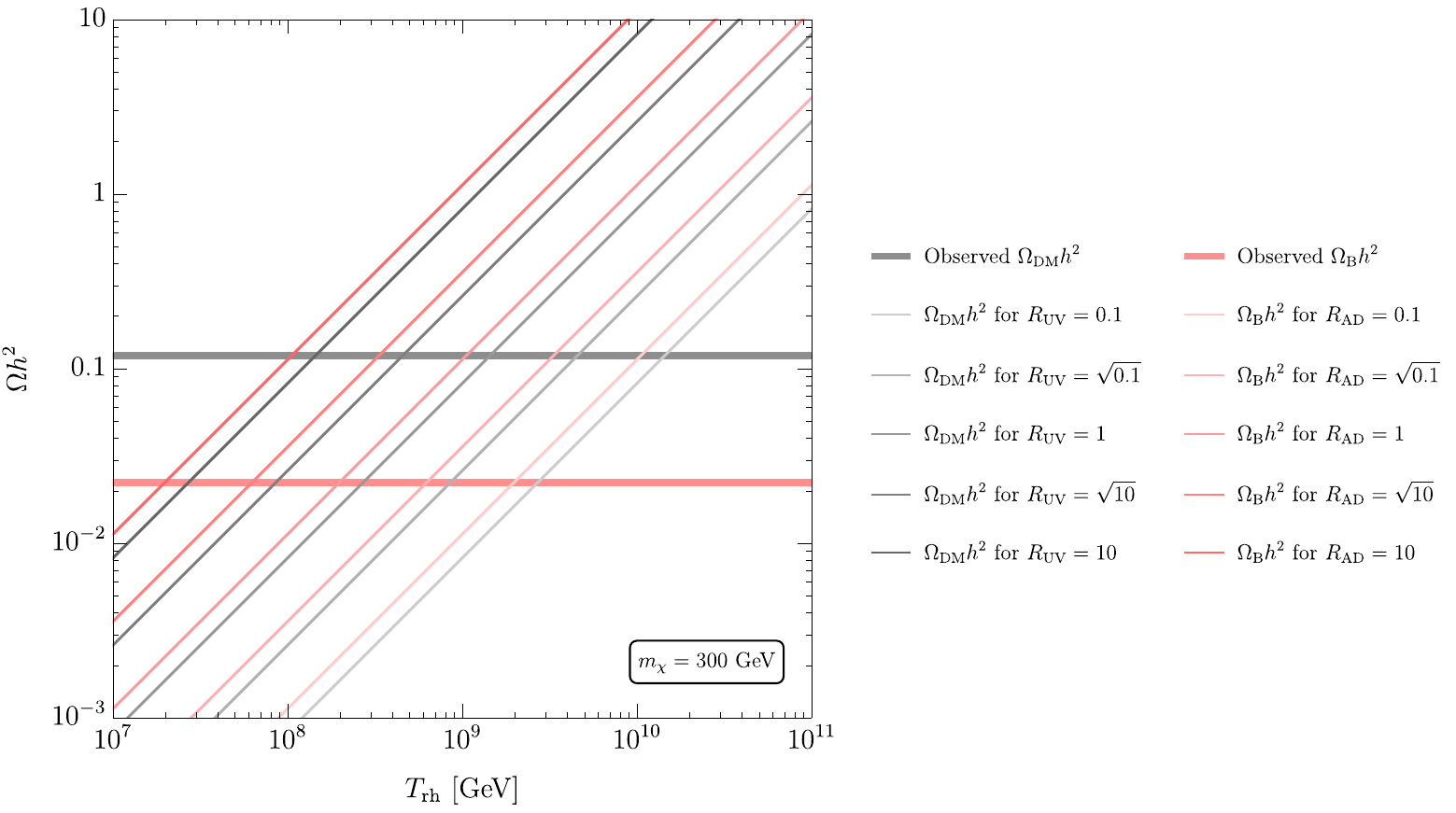}
\caption{The dark matter abundance from UV freeze-in and the baryon abundance from the Affleck-Dine mechanism in terms of the reheating temperature $\Trh$. Both abundances are proportional to $\Trh$, so the ratio between them does not depend on $\Trh$. With $\mathcal{O}(1)$ model parameters $\RUV$ and $R_{\rm AD}$  of eq.~(\ref{eq:nchi}) and (\ref{eq:Yphi}), the ratio is naturally $\mathcal{O}(1)$ as in \Eq{ratio}. We assume that the UV freeze-in portal and the operator that lifts the flat direction in the Affleck-Dine mechanism are both mass dimension five (i.e.~corresponding to~$n=4$ in the potential of eq.~(\ref{V})) with the cut-off scales both identified as the reduced Planck mass $\Mpl$.}
\label{fig:TrhOmega}
\end{figure*}

We next examine the ratio of the baryon density today $\Omega_\textrm{B}=\rho_\textrm{B}/\rho_c$ to the dark matter density today $\Omega_{\rm DM}=\rho_{\rm DM}/\rho_c$, where $\rho_\textrm{B}$, $\rho_{\rm DM}$, and $\rho_c$ are the baryon mass density, the dark matter mass density, and the critical density today, respectively. Assuming that the baryon asymmetry and dark matter relic abundance originate from the Affleck-Dine and UV freeze-in mechanisms, respectively, via leading higher-dimensional operators, we find
\bea 
\frac{\Omega_{\rm DM} }{\Omega_\textrm{B} } \label{ratio}
&= \frac{m_\chi Y_\chi}{m_N Y_\phi}
&=   \left(\frac{\cUV  m_{\chi}}{m_N}\right) \left(\frac{\RUV}{R_{\rm AD}}\right) ~,
\eea
where $m_\chi$ is the dark matter mass, and $m_N \approx 1 \GeV$ is the nucleon mass. We can see that the ratio between dark matter and baryon abundance does not depend on the reheating temperature but only on model parameters. If we choose $m_\chi \sim m_\textrm{soft} \sim \TeV$ and $\RUV \sim R_{\rm AD} \sim \mathcal{O}(1)$, we naturally have $\Omega_{\rm DM}/\Omega_\textrm{B} \sim \mathcal{O}(1)$.

We show $\Omega_{\rm DM} h^2$ from the UV freeze-in mechanism and $\Omega_\textrm{B} h^2$ from the Affleck-Dine mechanism in terms of $\Trh$ with different choices of model parameters $\RUV$ and $R_{\rm AD}$ in \Fig{fig:TrhOmega}. As standard, $h$ is the Hubble parameter today in the unit of $100~\textrm{km/s/Mpc}$. The figure clearly shows that the linear dependence on $\Trh$ and agreement between $\Omega_{\rm DM}$ and $\Omega_\textrm{B}$ as long as $\RUV$ and $R_{\rm AD}$ are $\mathcal{O}(1)$.

In what follows, we explore general conditions under which the parametric coincidence between the baryon asymmetry and dark matter abundance can be realized within specific model constructions, taking into account various phenomenological constraints.\footnote{We note in passing that there are alternative explanations for the coincidence of dark matter and baryon abundances involving the Affleck-Dine mechanism that appear in the literature \cite{Cheung:2011if,Rosa:2022sym}.} Supersymmetry provides a natural framework that accommodates viable UV freeze-in dark matter candidates and offers the flat directions required for the Affleck-Dine mechanism. We therefore focus on matter generation within SUSY and discuss its possible implications.

\section{UV Freeze-In DM in SUSY}
\label{Sec3}

Here, we discuss possible interactions between dark matter and the visible sector within the framework of supersymmetry, and then examine UV freeze-in gravitino dark matter in more detail.

\subsection{Higgs portal UV freeze-in}

Consider the Minimal Supersymmetric Standard Model (MSSM), extended by an additional gauge singlet chiral superfield 
${\mb X}= X + \sqrt{2}\theta \chi + \theta^2 F_X$. 
This introduces a new hidden sector Standard Model singlet fermion \(\chi\).
Enforcing a $Z_2$ parity  for $\mb X$, 
the renormalizable contribution to the superpotential is given by
\begin{equation}\label{HW}
\Delta W =  \frac{1}{2} \mu_X {\mb X}^2  + \frac{1}{\Lambda} {\mb X}^2 {\mb H_u} {\mb H_d},
\end{equation}
where, \(\mu_X\) provides the Majorana mass of dark matter candidate $\chi$. 
The superpotential of eq.~(\ref{HW}) implies the following dimension five operator in the Lagrangian  
\begin{equation} \label{eq:UVFI_LSP}
{\cal L}\supset
\frac{1}{\Lambda} \chi^2  H_u H_d + {\rm h.c.} 
\end{equation}
In addition to eq.~(\ref{eq:UVFI_LSP}), 
other dimension-five interactions arise, involving $X$ the scalar component of ${\mb X}$, and Higgsinos $\widetilde{H}_u$, $\widetilde{H}_d$, such as
\bea 
{\cal L}\supset
\frac{1}{\Lambda} \chi X (\widetilde{H}_u  H_d + H_u \widetilde{H}_d) + {\rm h.c.} 
\eea 
There also exist dimension-four operators suppressed by 
$\mu_X/\Lambda$ for $X$, but such operators are not relevant for dark matter production at high temperatures.

Famously, the MSSM  --and its extensions-- lead to (disastrously) fast proton decay unless certain operators are forbidden by R-parity (or a similar symmetry). If R-parity is exact, then the lightest supersymmetric particle (LSP) is stable. Thus if $\chi$ is the LSP, then it will be stable; interestingly, $X$ will also be stable as the lightest $Z_2$-parity state.  The dark matter density is then established by thermal production, as discussed in Sec. \ref{sec2:UVFI}, and the late time decays of the next-to-lightest supersymmetric particle (NLSP) into $X+\chi$. 

In the context of this class of models, we can identify the production coefficient $\RUV$, as given in eq.~(\ref{eq:UVFIthp}). For the abundance of $\chi$ and $X$ this coefficient will be  \cite{Elahi:2014fsa}
\beq
R_{\textrm{DM,HP}}
\simeq \frac{1}{16\pi^5}~.
\eeq

The scale $\Lambda$, which suppresses the dimension five operator, indicates the scale at which the operator is generated. This could be some intermediate scale, perhaps connected to mediation of SUSY breaking; another natural choice is $\Lambda=M_{\rm Pl}$.
In particular, in the case that the visible and hidden sectors are decoupled due to a preserved global symmetry, then gravitational interactions that violate these symmetries can provide a weak coupling of these two sectors. 
Moreover, supersymmetric models require at least one hidden sector, and the complexity of string compactifications commonly leads to multiple sectors that communicate only via non-renormalizable operators. 
Thus, if the  LSP resides in one of these hidden sectors, it is natural to think that it may be stable and that its abundance may be set by UV freeze-in via some higher dimension operators.

\subsection{Axino UV freeze-in}

Another clean and generic possibility for mass dimension five UV freeze-in arising in the context of supersymmetry come from superpotential terms involving field-strength supermultiplets 
\bea 
{\cal W}^a = \lambda^a + \theta (D^a - i \sigma^{\mu\nu} F^a_{\mu\nu}) 
-i\theta^2 \sigma^\mu D_\mu \bar\lambda^a,
\eea
where $F^a_{\mu\nu}$ represents the field strength of the gauge boson, and $\lambda^a$ is the gaugino, the fermionic superpartner of the gauge boson. We can introduce a dimension-five supersymmetric Lagrangian with the coupling between the chiral supermultiplet ${\mb X}$ and the field-strength superfields 
\bea\label{eq:UVFI_gaugino}
{\cal L} \supset\frac{1}{\Lambda}\int d^2\theta   {\mb X} {\cal W}^a {\cal W}^a + {\rm h.c.} 
\eea  
If R-parity is conserved, the fermion component $\chi$ of ${\mb X}$ can be stable if it is the LSP. The scalar superpartner $X$ is not stable and will decay to two gauge bosons. 
The relevant interactions for thermal production are 
\bea 
{\cal L}\supset \frac{\sqrt{2}}{\Lambda}
\left(i \chi \sigma^{\mu\nu} \lambda^a  G_{\mu\nu}^a 
 +  {\rm h.c.} + \sum_i \chi\lambda^a g_a \phi^\dagger_i T^a  \phi_i \right).
\eea
Here, $g_a$ is the gauge coupling, and  $\phi_i$ are the Standard Model charged scalar fields. 
At high temperatures, all gauginos and charged scalar particles are in thermal equilibrium. Therefore, the production of $\chi$ is the same as the usual UV freeze-in.

The most famous example of this class of models is the axino \cite{Covi:1999ty,Choi:2013lwa,Kim:1983ia,Rajagopal:1990yx,Goto:1991gq,Chun:1992zk,Covi:2001nw,Strumia:2010aa}. In this case ${\mb X}$ is identified as the axion supermultiplet, $\chi\equiv \tilde{a}$ is identified as the axino and $X\equiv a$ is the axion. 
The superpotential term of eq.~(\ref{eq:UVFI_gaugino}) is obtained at one-loop level from the anomalous global U(1) Peccei-Quinn (PQ) symmetry \cite{Peccei:1977hh}. This implies that $\Lambda$ is associated with the intermediate scale at which PQ breaking occurs, and one can make the matching
\begin{equation} 
\frac{1}{\Lambda} = \frac{g_a^2}{16\pi^2} \frac{1}{f_a},
\end{equation}
where $f_a$ is the axion decay constant. Moreover, from inspection of the existing literature, we can identify the production coefficient $\RUV$  for the axino states in the  Boltzmann equation of eq.~(\ref{eq:UVFIthp}) to be of the form \cite{Strumia:2010aa}
\bea
R_{\textrm{UV},\tilde{a}}
=  \frac{g_3^4}{256\pi^7 }\left(
 G(g_3) + 1.29 \times \frac{6g_3^2}{\pi^2}\right)~,
\eea
where $G$ is a numerical factor \cite{Rychkov:2007uq} with the following form
$G(g_3) = \frac{320}{\pi^2}  g_3^2\ln\left(1.2/g_3\right)$.
Evaluating with a characteristic guage coupling $g_3\simeq1$ gives $G(1)\simeq6$ and $R_{\chi,\tilde{a}} |_{g_3=1} \simeq9\times10^{-6}$. However, note that a full analysis should account for the running of $g_3$.

In the case that the axion $a$ is associated with the anomalous global U(1) that dynamically resolves the Strong CP problem, then $f_a$ is typically expected to be at some intermediate scale $m_Z\ll f_a\ll M_{\rm Pl}$.   If this axion is unrelated to the Strong CP problem $\Lambda$ has an even greater degree of freedom. Moreover, $f_a={\cal O}(0.01-0.1)M_{\rm Pl}$ is well motivated in the case of string axions \cite{Svrcek:2006yi,Arvanitaki:2009fg}. 

\subsection{UV Freeze-in gravitino dark matter}
\label{1234}

Another prime candidate for SUSY UV freeze-in is the gravitino, the gauge fermion superpartner of the graviton with spin $3/2$.  The gravitino includes the  {\it goldstino} as a spin $1/2$ component of a Nambu-Goldstone fermion that arises from the spontaneous breaking of rigid supersymmetry. At high temperatures ($m_{3/2}\gg T$, where $m_{3/2}$ is the gravitino mass), the thermal production of the longitudinal component of gravitino can be described by goldstino production \cite{Casalbuoni:1988kv}.
The qualitative feature of the interaction strength between the goldstino and Standard Model particles can be easily understood from eq.~(\ref{eq:UVFI_gaugino}). 
When the spontaneous breaking of SUSY is dominantly given by the nonzero expectation value $F_X$ of  $\mb X$, the fermion component of the superfield $\mb X$ is the goldstino $\chi$. It is subsequently absorbed by the gravitino, which becomes massive with a mass $m_{3/2} = F_X/(\sqrt{3}M_{\rm Pl})$.  

According to eq.~(\ref{eq:UVFI_gaugino}), the gaugino mass is given by 
\bea 
m_{\lambda^a} = \frac{F_X}{\Lambda}. 
\eea
Thus, we can express the inverse of $\Lambda$ as
\bea \label{eq:goldstino}
\frac{1}{\Lambda}= \frac{m_{\lambda^a}}{F_X} = \frac{1}{M_{\rm Pl}} \frac{m_{\lambda^a}}{\sqrt{3}m_{3/2}}.
\eea
This indicates that for $m_\lambda \gg m_{3/2}$ the goldstino coupling to the Standard Model particles has an enhancement factor of the gaugino-to-gravitino mass ratio, in addition to the Planck mass suppression. As a result, goldstino production dominates the thermal production of gravitino.

More generally, supergravity restricts the gravitino $\psi_\mu$ interactions to the MSSM particles to have the following form
\bea 
{\cal L} =& - \frac{1}{M_{\rm Pl}}\left(\frac{i}{2} \psi_\mu \sigma^{\nu\rho}\sigma^\mu \bar\lambda^a F^a_{\nu\rho} + {\rm h.c.}  \right. \\ 
& \hskip 1.3cm \left.  + \frac{i}{\sqrt{2}}(D_\mu \phi_i)^\dagger \psi_\nu \sigma^{\mu}\bar\sigma^\nu \psi_i + {\rm h.c.}  \right) .
\eea 
The interaction of the goldstino component can be derived through the substitution $\psi_\mu\to\sqrt{(2/3)}\frac{1}{m_{3/2}}(\partial_\mu\chi)$. This leads to the same interaction strength as given in eq.~(\ref{eq:goldstino}) upon the integration by parts and application of the equations of motion for gauginos.

In gauge mediated supersymmetry breaking, the gravitino mass is suppressed by a factor of $\Lambda/M_{\rm Pl}$ relative to Standard Model superpartners. As a result, the gravitino is generically the LSP when supersymmetry breaking is communicated via gauge mediation.
Identifying the LSP for gravity-mediated supersymmetry breaking is more complicated since all superpartners have masses of order the gravitino mass $m_{3/2}\sim F_X/M_{\rm Pl}$.
Notably, it was shown by Kersten \& Lebedev \cite{Kersten:2009qk}  that for gravity-mediated supersymmetry breaking the gravitino will be the LSP if at the GUT scale 
\beq\label{cond}
\left.  \frac{m_{1/2}}{m_{3/2}}\right|_{\mu=M_{\rm GUT}}\gtrsim2.6 ,  
\eeq
where $m_{1/2}$ is the unified gaugino mass $m_{\lambda^a}$ at the GUT scale. 
This GUT boundary condition is determined by the form of the gauge kinetic function and the K\"ahler potential. 
The inequality of eq.~(\ref{cond}) can be readily satisfied in string-motivated setups \cite{Kersten:2009qk}. Furthermore, in this case, either the neutralino or stau will typically be the next-to-lightest superpartner (NLSP).

There have been many studies of the thermal production of gravitinos \cite{Rychkov:2007uq,Moroi:1993mb,Pradler:2006qh,Cyburt:2002uv,Bolz,Moroi:1995fs,Kaneta:2023uwi,Ellis:1984eq,Ellis:1995mr,Garcia:2017tuj,Bernal:2019mhf,Ellis:2015jpg,Cheung:2011nn,Covi:2009bk,Eberl:2020fml}. Here in examining gravitino production rate we follow the recent work of Eberl, Gialamas \& Spanos \cite{Eberl:2020fml} which includes the one-loop thermal self-energy, and claims to correct errors in the literature.\footnote{Another contemporary treatment of the gravitino production rate leads to an $\mathcal{O}(10\%)$ difference in the yield \cite{Kaneta:2023uwi}.} 
The Boltzmann equation for the UV freeze-in production eq.~(\ref{eq:UVFIthp}) 
can be specified for the gravitino in terms of  
$R_{\textrm{UV},3/2}$, which is found to be \cite{Eberl:2020fml}
\bea\label{r32}
 R_{\textrm{UV},3/2}=&
\frac{3\zeta(3)}{16\pi^3}
\sum_a c_a g_a^2\left(1+\frac{m_{\lambda^a}^2}{3 m_{3/2}^2}\right){\rm ln}\left(\frac{k_a}{g_a}\right)\\
&+
72C\lambda_t^2\left(1+\frac{A_t^2}{3m_{3/2}^2}\right)~,
\eea
where  $c_a=(41.9,~68.2,~21.1)$  and $k_a\simeq(0.8,~1.0,~6.9)$ are constants labelled by the corresponding index $a=1,2,3$ associated to the gauge couplings $g_a$. 

The final term in eq.~(\ref{r32}) corresponds to $\tilde{t}\tilde{t}\rightarrow \tilde{H}\tilde{\psi}$ and involves  the top Yukawa $\lambda_t$, the top-stop A-term $A_t$ and a constant $C\approx2.6\times10^{-4}$.
When the $A_t$-term can be neglected the characteristic size for $R_{\chi,3/2}$ is
\beq\label{24}
 R_{\textrm{UV},3/2}\sim
0.4 \left(1+\frac{m_{1/2}^2}{3m_{3/2}^2}\right)~,
\eeq
where we have assumed a universal mass for the gauginos $m_{1/2}\simeq m_{\lambda^a}$ and evaluate the couplings $g_a$ at their weak scale values (using instead the coupling at the GUT scale $\alpha_{\rm GUT}=g_{\rm GUT}^2/4\pi=1/24$  halves the estimate of $R_{\textrm{UV},3/2}$).

For a gravitino LSP whose abundance is entirely determined by thermal production, the dark matter relic abundance is related to the freeze-in yield $Y_{3/2}$ via 
\beq\label{ooo}
\Omega_{3/2}h^2  = \frac{45   \zeta(3)}{2\pi^4 g_S^*} \frac{m_{3/2}Y_{3/2}(T_0) s(T_0)h^2}{\rho_c}~,
\eeq
where $T_0\approx2.348\times10^{-13}$ GeV is the present-day temperature of the cosmic microwave background. 
It follows that the gravitino freeze-in yield is  given by 
\beq\label{Y32}
Y_{3/2}=\frac{n_{3/2}}{s}
\simeq
\frac{R_{\textrm{UV},3/2}(T_{\rm rh})}{H(T_{\rm rh})s(T_{\rm rh})}
\frac{g_{s}^*(T)}{g^*_{s}(T_{\rm rh})}~.
\eeq

To evaluate $R_{\textrm{UV},3/2}(T_{\rm rh})$ in eq.~(\ref{Y32}) one must run the gauge couplings into the UV. In  \cite{Eberl:2020fml} they take the couplings to be $g_a(T_{\rm rh})=g_{\rm GUT}$. Implementing the one-loop renormalization group evolution for the gauge couplings and evaluating each at $g_a(\Trh)$ is straightforward.

Putting eq.~(\ref{ooo}) and eq.~(\ref{Y32}) together and evaluating the numerical prefactor one obtains
\beq
\Omega_{3/2}h^2=1.63\times10^{24}
\Big(\frac{m_{3/2} R_{\textrm{UV},3/2} T_{\rm rh}}{M_{\rm Pl}^2}\Big)~.
\label{o32}
\eeq

The form of eq.~(\ref{Y32}) assumes that inflaton decay is instantaneous. More realistically, decays should exponentially deplete the inflaton energy density. From comparison with \cite{Garcia:2017tuj}, it is anticipated that a more careful treatment of inflaton decay leads to an $\mathcal{O}(10\%)$ correction \cite{Rychkov:2007uq,Eberl:2020fml}. An analytic argument in  Appendix \ref{ApB} also supports this numerical study that such non-instantaneous reheating effects should be small. Thus, here, we will also neglect these effects.

We close by noting that NSLP decay to gravitinos can lead to major complications if they contribute significantly to $\Omega_{3/2}h^2$. If R-parity is conserved one anticipates that its contribution to the gravitino abundance from NLPS decays to be given by
\beq \label{nlsp}
\Delta\Omega_{3/2} h^2 = \frac{m_{3/2}}{m_{\text{NLSP}}} \Omega_{\text{NLSP}} h^2~.
\eeq
For gauge mediation one anticipates that $m_{3/2}\ll m_{\text{NLSP}}$ in which case the contribution  $\Delta\Omega_{3/2} h^2$ is naturally suppressed. For freeze-in to set the gravitino abundance, one requires $\Delta\Omega_{3/2} h^2 \ll \Omega_{3/2} h^2 $, otherwise the link to $\Trh$, and thus Affleck-Dine baryogenesis, is spoiled.
 In Section \ref{Sec6} we identify two scenarios in which this is not a concern, namely a RH sneutrino NLSP and a scenario with R-partity violation. Appendix  \ref{NLSP-Decays} also provides a discussion on implications for alternative NLSP candidates, such as the stau or neutralino.

\section{Affleck-Dine Baryogenesis in SUSY}
\label{Sec4}

In supersymmetric models, the renormalizable scalar potential is determined by sums involving the squares of  F-terms and D-terms. As a result, for certain `flat directions' in the field space, the scalar potential may be exactly zero. The seminal paper of Dine, Randall, \& Thomas \cite{Dine:1996} highlighted that MSSM flat directions with non-zero $B-L$ are ideal for implementing the Affleck-Dine mechanism. These flat directions are lifted by soft SUSY-breaking terms and non-renormalizable terms, which generate a non-vanishing scalar potential for the field $\phi$ parameterizing the flat direction.

\subsection{Asymmetry generation}
\label{2A}

All forms of supersymmetry breaking will weakly lift any flat directions. The general form of this lifting term can be parameterized by the non-renormalizable operator in superpotential 
\bea 
\mathcal{O}_n= \frac{\phi^n}{\Lambda^{n-3}}~.
\eea 
If gravity alone mediates SUSY breaking this introduces a soft term $m_{\phi}^2 |\phi|^2$, where $\phi$ is the scalar component of the field parameterizing the flat direction, and $m_\phi \simeq m_{3/2}$ coincides with the gravity-mediated SUSY breaking scale.
In the case of gravity-mediated supersymmetry breaking for which  scalar potential is lifted by the leading non-renormalizable operator $\mathcal{O}_n$, the resulting  scalar potential can be expressed as follows \cite{Enqvist:1997si,Enqvist:1999mv}
\bea\label{V}
V\simeq  & ~(m_{\phi}^2-cH^2)|\phi|^2 \\
&+\frac{1}{\Lambda^{2n-6}}|\phi|^{2n-2}
+\left(\frac{A \phi^n}{n\Lambda^{n-3}}+{\rm h.c.}\right)~,
\eea
The Hubble parameter $H$ accounts for SUSY breaking due to the inflaton energy density. The constant $c$ is related to the coupling between the inflaton and the $\phi$ field and should be positive and $\mathcal{O}(1)$ to ensure that $\phi$ develops a large expectation value during inflation \cite{Dine:1996}. 
In the Affleck-Dine mechanism, the field associated with the flat direction carries a conserved global charge  $Q$ (this is typically taken to $B$ or $L$). 

At the end of inflation, the $\phi$ field will be at a large field value provided $cH^2>0$. It subsequently evolves to form a coherently oscillating field at $H_{\rm osc} =  m_\phi/\gamma$ where $\gamma$ parameterises $\mathcal{O}(1)$ factors. During this period of oscillations, the symmetry-violating A-term $A\phi^n$ causes the $\phi$ condensate to develop a global charge asymmetry (see e.g.~\cite{Fujii:2002kr} for details). The resulting $\phi$-asymmetry can be expressed as follows 
\beq \label{eta}
Y_\phi^{(n)}=\delta\left(\frac{m_{3/2}T_{\rm rh}}{M_{\rm Pl}^2m_\phi^2} \right)  (m_\phi \Lambda^{n-3})^{\frac{2}{n-2}}~,
\eeq
where $\delta$ collects all the other various $\mathcal{O}(1)$ factors 
\beq
\delta=\frac{n-2}{6(n-3)}|a_m| \beta\gamma^{\frac{2(n-3)}{n-2}}\sin\theta~.
\label{delta}
\eeq
Here $\beta$ is the global charge (i.e.~baryon/lepton number) of $\phi$, we parameterize the A-term $A = a_m m_{3/2}$, and the phases enter via 
\beq
\theta={\rm arg}[a_m]+{\rm arg}[\phi]~.
\eeq
 Without tunings, it is natural to expect the value of $\delta$ to lie between 0.01 and 1.
Thus, for a dimension 4 operator (such as the ${\mb H_u}{\mb L}$ flat direction \cite{Murayama:1993em}) with $\Lambda=M_{\rm Pl}$ the asymmetry is 
\beq\label{eta4}
Y^{(4)}_\phi= \frac{|a_m|\beta\gamma\sin\theta}{3}\left(\frac{m_{3/2}T_{\rm rh}}{m_\phi M_{\rm Pl}} \right).
\eeq

Notably, the Affleck-Dine mechanism is parametrically different in the presence of gauge-mediated SUSY-breaking. 
 For gauge-mediation scenarios, hidden sector SUSY breaking is communicated to the visible sector through messenger fields (with mass $M$), which are non-trivial representations of the Standard Model gauge groups. This results in soft-breaking terms, which are distinct from the case of gravity-mediated SUSY breaking. In particular, the soft mass scale is parametrically
\beq
m_{\rm soft}\sim \frac{F_X}{M} \gg m_{3/2}\sim\frac{F_X}{\mpl}~.
\label{soft}
\eeq
In the case that  $M\ll M_{\rm Pl}$, the gravitino can be significantly lighter than the spectrum of Standard Model superpartners, making the gravitino generically the LSP. The introduction of gauge mediation also impacts the Affleck-Dine mechanism.
For SUSY breaking mediation mechanisms other than gravity mediation, the  $\phi$ potential will receive a corresponding soft breaking mass (in addition to the model-independent gravity mediated contribution), implying a contribution to the $\phi$ potential of the form  \cite{Doddato:2012ja,Kusenko:1997si,Kasuya:2001hg,Kasuya:2001tp,deGouvea:1997afu}\footnote{There can also be thermal contributions, in particular thermal mass terms of the form $\sum_{f_k |\phi| < T}  f^2_k T^2 |\phi|^2$ \cite{Asaka:2000nb,Allahverdi:2000zd,Anisimov:2000wx}. These contributions can be neglected for small couplings $f^2_k$, as we discuss for the specific models we study in Section \ref{Sec6}. \label{foot2}}   
\bea
\Delta V(\Phi) \sim\begin{cases}
m_{\rm soft}^2 M^2 \ln \left(1 + \frac{|\phi|^2}{M^2}\right)   & \qquad   |\Phi|>M \\ 
F_X^2 \ln \left(\frac{|\phi|^2}{M^2}\right)    & \qquad |\Phi|<M
\end{cases}
\label{Vgauge}
\eea
This potential term supplements the model-independent gravity-mediated terms of eq.~(\ref{V}). For extremely large field values, the gravity mediated contribution dominates \cite{deGouvea:1997afu} and the asymmetry is unchanged from eq.~(\ref{eta}). In contrast, if the contributions due to gauge mediation dominate (for small or modest initial field values), then the asymmetry generated differs by a factor of $m_{\rm soft}/m_{3/2}$ (see e.g.~\cite{Enqvist:2003gh})
\beq
Y^{(n)}_\phi \sim  R_{\rm AD}^{(n)}\left(\frac{\Trh}{M_{\rm Pl}}\right)
\eeq 
with 
\beq
R_{\rm AD}^{(n)} = \frac{\beta(n-2)}{6(n-3)}\Big(\frac{m_{\rm soft}}{m_\phi}\Big) \Big(\frac{\Lambda}{M_{\rm Pl}}\Big)^\frac{2(n-3)}{n-2}
\Big(\frac{M_{\rm Pl}}{m_\phi}\Big)^\frac{n-4}{n-2},
\eeq
where we have dropped $\mathcal{O}(1)$ factors, such as CP phases, and we take $H_{\rm osc} \simeq  m_\phi$. 

Importantly, in many cases, the Affleck-Dine condensate is not the lowest energy configuration. As a result, spatial perturbations may cause the condensate to fragment into non-topological solitons, called Q-balls \cite{Coleman:1985ki,Kusenko:1997ad,Kusenko:1997zq,Enqvist:1999mv}. These Q-balls are meta-stable, charge $Q$ solitons associated to scalar fields with a spontaneously broken global U(1)  \cite{Coleman:1985ki} and can have significant implications for baryon asymmetry generation, as we discuss next.

\subsection{Avoiding Q-ball formation}
\label{2B}

The link between Affleck-Dine and UV freeze-in can be broken if other processes influence the late-time dark matter abundance or baryon asymmetry. If the Affleck-Dine condensate fragments into Q-balls, this will generally lead to one or both of $\Omega_\textrm{B}$ and $\Omega_{\rm DM}$ no longer being dominantly determined by the reheat temperature $T_{\rm rh}$. Such a change spoils the elegant connection established in eq.~(\ref{ratio}). Accordingly, below we discuss how to avoid Q-ball formation, which is the simplest manner to circumvent this issue.\footnote{The connection between  $\Omega_{\rm DM}$ and $\Omega_\textrm{B}$ can potentially be maintained if Q-balls do form but then rapidly evaporate. Fast Q-ball decay can occur for low $T_{\rm rh}$ and/or high $m_\phi$ since the lifetime is controlled by $\tau\propto Q$ and $Q\propto \sqrt{K}m_\phi/T_{\rm rh}$ \cite{Cohen:1986ct,Enqvist:2003gh,Enqvist:1997si}. One would require the gravitino production to be highly subdominant. This seems less clean and we do not explore it here.} 

If the scalar potential along the flat direction is shallower than $\phi^2$, the $\phi$ condensate experiences negative pressure, which leads to the growth of inflation-induced perturbations, and the fragmentation of the inflaton into Q-balls \cite{Turner:1983he,Enqvist:1997si,Enqvist:2003gh}. In the case of gravity-mediated supersymmetry breaking, the formation of Q-balls is determined by the one-loop correction to the quartic term
\bea\label{Vgrav}
\Delta V\simeq &~ (m_{\phi}^2-cH^2)  K \log\left(\frac{|\phi|^2}{M_{\rm Pl}^2}\right) |\phi|^2~.
\eea
The factor $K$ can be calculated for a given flat direction and is a function of energy scale $K=K(\mu)$.
Notably, the sign of $K$ determines whether Q-balls form. 
Specifically, for $K > 0$, the $\phi$ potential grows faster than quadratic,  the $\phi$-condensate remains stable, and Q-balls do not form. Whereas in the converse case $K<0$, the potential grows slower than quadratic, leading to Q-ball formation.
Note that gaugino loops contribute negatively to $K$, while Yukawa couplings contribute positively.

Condensate collapse and Q-ball formation are generically observed for almost all of the MSSM flat directions \cite{Enqvist:2000gq}. However, for the leptonic flat directions $K$ is generically positive \cite{Enqvist:2000gq,Enqvist:1997si,Fujii:2001dn}, since the leptonic flat directions do not couple to the gluino, which drives the squark flat directions to negative $K$.  In special parts of parameter space, squark flat directions can have $K>0$; however, this requires the flat direction to be more strongly aligned with the third-generation squarks, e.g.~$Q_3Q_3QL$ \cite{Enqvist:2000gq}.

Another prospect is the RH~sneutrino $\phi=\widetilde{\nu}_R$ \cite{Casas:1997gx} with, for example, a scalar potential of the form
\bea\label{VRH}
V\simeq  & ~(m_{\phi}^2-cH^2)|\widetilde{\nu}_R|^2 
+\frac{1}{M_{\rm Pl}^{2}}|\widetilde{\nu}_R|^6
+\left(\frac{A \widetilde{\nu}_R^4}{4M_{\rm Pl}}+{\rm h.c.}\right).
\eea
The one-loop running for $\phi=\widetilde{\nu}_R$  is such that $K$ is strictly positive since the RH sneutrino has no gauge couplings.
We will return to discuss the details of this sneutrino model in Section \ref{Sec6}.

In gauge mediation Q-ball physics can be altered since the scalar potential is different. For gauge mediated SUSY breaking the scalar potential picks up additional terms of the form of eq.~(\ref{Vgauge}). In the case that $|\phi|/M\gg1$, then the potential is dominated by gravity-mediated contribution, and the analysis is unchanged compared to earlier subsections. Conversely, for $|\phi|/M\lesssim1$ gauge meditation introduces the slow-growing logarithmic potential, and Q-balls will generically form (regardless of the sign of $K$) \cite{Kasuya:2001hg} (see also \cite{Doddato:2012ja,Kusenko:1997si,Kasuya:2001hg}). 
Thus for gauge mediation scenarios with $K>0$,  Q-balls will  form  only while the field value of the Affleck-Dine field $\phi$ lies in a window~\cite{Enqvist:2003gh}
\beq
m_\phi <\phi< \sqrt{2} \left(\frac{m_{\rm soft}^2}{m_{3/2}}\right)~,
\eeq
 where the $\phi$ mass $m_\phi$ corresponds to the temperature at which the condensate transfers its energy density to light degrees of freedom. In certain models, this window may be closed.

For gauge-mediated SUSY breaking, the simplest scenario that avoids Q-ball formation is the identification of the RH sneutrino with the Affleck-Dine field $\phi=\widetilde{\nu}_R$. Since $\widetilde{\nu}_R$ is a singlet under the Standard Model gauge group it receives highly suppressed gauge mediated contributions to its soft breaking mass. This is especially true if the RH sneutrinos have small Yukawa couplings.  In Section \ref{Sec6} we will explore a second scenario (based on R-partity violation) in which Q-balls are absent due to the small couplings of the Affleck-Dine field. 

Another interesting possibility we highlight in passing for avoiding Q-balls in gauge mediation is to use the ${\mb H_u}{\mb L}$ flat direction $\phi/\sqrt{2}= H_u=L $ with a gauged U(1)${}_{B-L}$ symmetry, as proposed in \cite{Fujii:2001dn}. This scenario is quite distinct from other models since the scale $\Lambda$  at which non-renormalizable operators in the Affleck-Dine potential arise is related to the scale of U(1)${}_{B-L}$  breaking (rather than $M_{\rm Pl}$). The scale $\Lambda$ can be high, but should be below the messenger scale  $\Lambda<M$. 
In this case,  the scale $\Lambda$ is also related the RH neutrino mass scale \cite{Moroi:1999uc} and thus the light neutrino masses with $m_\nu\sim \langle H_u\rangle^2/M$.  This typically implies the lightest neutrino $m_1$ is hierarchically lighter than the other two species \cite{Asaka:2000nb}.

\section{Models}
\label{Sec6}
We next identify two specific examples based on supersymmetric models, in which the gravitino ubiquitous in SUSY frameworks is considered the dark matter candidate as the LSP. In these scenarios, the observed ratio $\Omega_{\rm DM}/\Omega_\textrm{B}$ becomes insensitive to $\Trh$ due to the parametric coincidence between Affleck-Dine baryogenesis and UV freeze-in. When the gravitino constitutes dark matter, a natural concern arises: supersymmetric partners of Standard Model particles, particularly the NLSP, must eventually decay into gravitinos. However, the NLSP typically has a long lifetime, which can severely affect Big Bang Nucleosynthesis (BBN) and may lead to an overproduction of gravitinos. We investigate two concrete models that naturally address these issues, where the associated interactions and sectors responsible for these decays also play an essential role in generating the baryon asymmetry. Specifically, we examine the cases in which the Affleck-Dine field corresponds to either a right-handed sneutrino or an R-parity violating flat direction. These models significantly relax the constraints from NLSP decays; to make this point clear, we begin by discussing the bounds arising from BBN observations.

\subsection{Gravitino dark matter and constraints}

Since the lifetime for an NLSP decaying to gravitinos is typically long, this commonly leads to energy injection during BBN, at which the temperature of the Universe is around 1 MeV. Observations of the relative abundances of primordial elements strongly favour standard cosmology during BBN, see e.g.~\cite{Kawasaki:2004qu,Sarkar:1995dd,Jedamzik:2006xz}. These studies strongly exclude decays of cosmological populations of states to baryonic matter. Since decays to gravitinos are Planck-suppressed, the lifetimes of the NLSP tend to be sufficiently long that this is a concern. Moreover, with the possible exception of resonantly annihilating NLSPs one expects the NLSP to have a sizeable cosmological abundance prior to decays. Thus, in this section, we consider the constraints that arise from BBN considerations on the NLSP candidates identified above.

The constraints from BBN depend on three factors, the abundance of the NLSP at the time of decays $Y_{\rm NLSP}$, the lifetime of the NLSP $\tau_{\rm NLSP}$, and the dominant decay channel (specifically if it is hadronic, electromagnetic, or neither). If the NLSP decays in a manner that has negligible hadronic and electromagnetic branching fractions (as is typically the case for the sneutrino), then the constraints are largely circumvented. Additionally, if $\tau_{\rm NLSP}<1$s then the NLSPs will decay prior to BBN. The BBN limits on NLSP decays to gravitino LSPs have been studied carefully for the stau \cite{Steffen:2006hw} and neutralino \cite{Covi:2009bk} (see also see also e.g.~\cite{Kawasaki:2004qu,Kanzaki:2006hm,Kawasaki:2008qe,Steffen:2006hw}). For a given NLSP candidate, one can constrain the abundance $Y_{\rm NLSP}$ as a function of  $\tau_{\rm NLSP}$, both of which are related to model parameters such as the NLSP mass and couplings, and the rest of the superpartner spectrum.

To be concrete, we consider the case of a stau NLSP, drawing on the ``conservative'' bounds derived by Pradler \& Steffen \cite{Steffen:2006hw}. 
 We first note that for  $\tau_{\tilde{\tau}}\lesssim100$s there are no exclusion limits. 
The decay rate of the scalar superpartner of a Standard Model fermion, such as the stau, to the gravitino is \cite{Moroi:1995fs}
\bea 
\Gamma(\tilde{\tau} \to \psi_{3/2}+\tau) = \frac{1}{48\pi} \frac{m_{\tilde{\tau} }^5}{m_{3/2}^2 M_{\rm Pl}^2} \left[1-\left(\frac{m_{3/2}}{m_{\tilde{\tau} }}\right)^2 \right]^2.
\eea 
For the case  $m_{3/2}\ll m_{\rm NLSP}$ the lifetime is
\beq\label{life}
\tau_{\widetilde{\tau}}\simeq 60~{\rm s}
\left(\frac{m_{3/2} }{10~{\rm GeV}}\right)^2
\left(\frac{1~{\rm TeV}}{m_{\tilde{\tau} }}\right)^5.
\eeq
Requiring $\tau_{\tilde{\tau}}\lesssim100$s gives the upper value for $m_{3/2}$ (for a given $m_{\tilde{\tau}}$). In this case the BBN limits (conservative and aggressive) \cite{Steffen:2006hw} are not constraining regardless of $Y_{\tilde{\tau}}$. Such short-lived stau NLSPs require a gravitino mass
\beq
m_{3/2}\lesssim 13~{\rm GeV}\left(\frac{m_{\tilde{\tau} }}{1~{\rm TeV}}\right)^5~.
\eeq

More generally, in  \cite{Steffen:2006hw}  it was found that for a $\tilde{\tau}$ NLSP  the model is safe provided $\tau_{\tilde{\tau}}\lesssim10^7$s and $\Omega_{\tilde{\tau}}/ \Omega_{\rm DM}\lesssim0.1$, this corresponds to 
\beq
m_{\tilde{\tau}}Y_{\tilde{\tau}}\lesssim6\times10^{-11}~{\rm GeV}.
\eeq
Let us consider the case that $m_{\tilde{\tau}}\sim m_{3/2}$ in which case the lifetime is $\tau_{\widetilde{\tau}}\sim 6\times10^5~{\rm s}
\times (m_{\tilde{\tau}}/1~{\rm TeV})^3$ and the freeze-out yield away from resonance is \cite{Okada:2007na}
\beq
m_{\tilde{\tau}}Y_{\tilde{\tau}}\sim10^{-9}~{\rm GeV}\left(\frac{m_{\tilde{\tau}}}{1~{\rm TeV}}\right)^2~.
\eeq
This is in conflict with the conservative limit of \cite{Steffen:2006hw} for TeV stau NLSP. Evading the conservative limits requires very light staus with $m_{\tilde{\tau}}\lesssim250$ GeV, and this is entirely excluded for the aggressive BBN assumptions of \cite{Steffen:2006hw}. 

Notably, resonant enhancement can commonly lead to a reduction of the $Y_{\tilde{\tau}}$ by a factor of $\mathcal{O}(10^{-2}-10^{-3})$ for $m_{\tilde\tau}\approx m_H/2$, see  \cite{Pradler:2008qc,Heisig:2013rya}. Thus, not only can resonant annihilation be helpful in avoiding NLSP decays from being the dominant source of gravitino production, it also aids in evading BBN constraints for TeV scale stau NLSP. 

We conclude that in the case of gauge mediation, the $\tilde\tau$ are a viable NLSP candidate provided $m_{3/2}\ll m_{\tilde\tau}$. While for vanilla gravity mediation with $m_{\tilde{\tau}}\sim m_{3/2}$ resonant annihilation is required to satisfy the (conservative) BBN bounds for the case that the stau NLSP decays only to gravitinos. Moreover, neutralino NLSP leads to a similar conclusion (see  e.g.~\cite{Covi:2009bk}).

We have identified two interesting alternative scenarios that avoid the need for (tuned) resonant NLSP annihilations. The first is the case of sneutrino NLSP in which case decays to gravitinos are neither hadronic nor electromagnetic, leading only to late time neutrino production which is hard to constrain. The second case is NLSPs decay via R-parity violating operators which can lead to the branching fraction of NLSP to gravitinos to be small. We will discuss both of these possibilities below.

\subsection{Sneutrino Affleck-Dine field}

\begin{figure*}[t]
\centering
\includegraphics[width=0.65\textwidth]{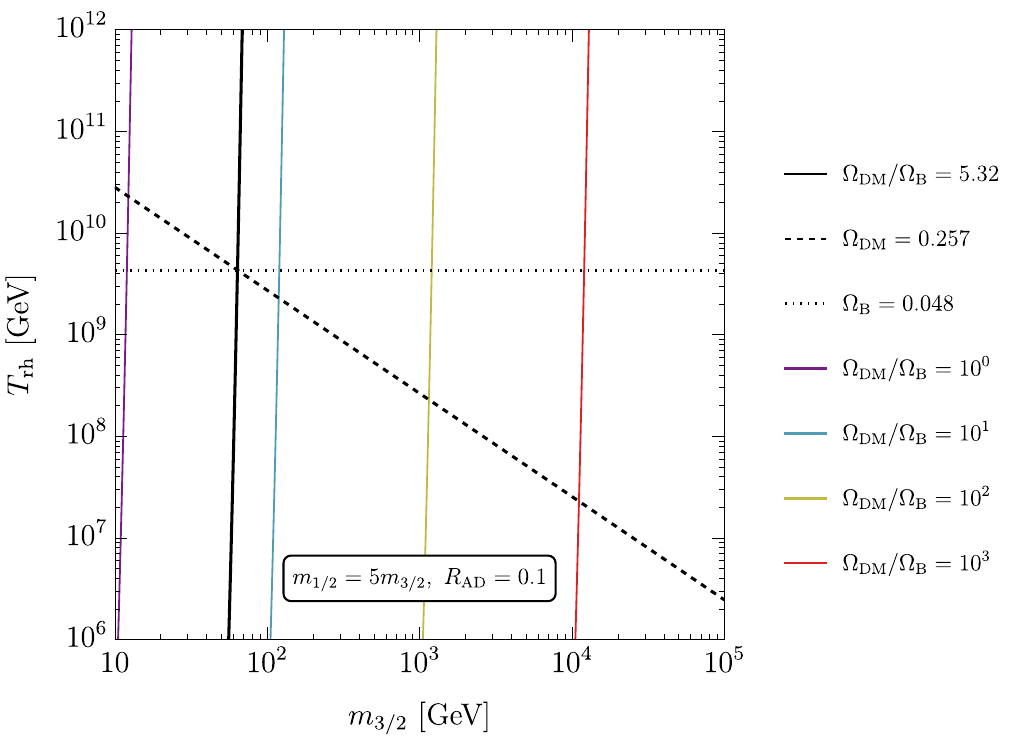}
\caption{
The plot shows the impact of varying the gravitino LSP mass $m_{3/2}$ and the reheat temperature $\Trh$ on the ratio of $\Omega_{\rm DM}/\Omega_\textrm{B}$. It is assumed that the baryon abundance $\Omega_\textrm{B}$ arises from to Affleck-Dine baryogenesis with $\phi=\widetilde{\nu}_R$ and relic abundance of gravitino LSP  is due to UV freeze-in $\Omega_{\rm DM}$. We assume that NLSP decays to gravitinos lead to a negligible contribution to $\Omega_{\rm DM}$, as is the case for our benchmark model with $\sin\theta_{\tilde \nu} = 0.1$ and $m_{\tilde L} = 5 m_{\tilde \nu_1}$ (see main text). Q-balls production will also be generically absent. As a result, the ratio of $\Omega_{\rm DM}/\Omega_\textrm{B}$ is constant with $\Trh$ as discussed in Section \ref{Sec2}. The near-vertical lines indicate the $T_{\rm RH}$-independence of the ratio $\Omega_{\rm DM}/\Omega_\textrm{B}$, with the slight variation arising due to mild renormalisation effects. }
\label{fig:Trhm32}
\end{figure*}

The case in which the sneutrino is identified as both the NLSP and the Affleck-Dine field $\phi$ has a number of distinct advantages in realizing the coincidence between $\Omega_\textrm{B}$ and $\Omega_{\rm DM}$. Specifically, $\phi=\widetilde{\nu}_R$ is a natural flat direction (cf.~eq.~(\ref{VRH})), Q-balls are typically absent. Moreover, the sneutrino soft mass can naturally lie below that of other MSSM superpartners, making it a generic NLSP candidate in a model with a gravitino LSP. Consequently, BBN constraints on NLSP decays are generally weak. In addition, thermal corrections \cite{Asaka:2000nb} to the quadratic term $|\phi|^2$ (cf.~Footnote \ref{foot2}) can be safely neglected, provided that the Yukawa couplings in $\Delta W_\nu = y_\nu {\mb L} {\mb H}_u{\mb \nu}_R$ is sufficiently small, a condition that is typically well justified.
Given these advantages, we now focus on this ideal subclass of models with $\phi = \widetilde{\nu}_R$.

The asymmetry in $\phi$ is generated via the Affleck-Dine mechanism.   To transfer this asymmetry to the baryons and leptons, we introduce a sizable $A$-term associated with sneutrinos-Higgs interactions \cite{Arkani-Hamed:2000oup}:
\bea \label{sneutrinoInt}
\mathcal{L} \supset A_\nu \tilde L H_u \widetilde{\nu}_R + {\rm h.c.}
\eea
When the $A$-term is generated from higher-dimensional operators in the K\"ahler potential,  it is expected to be of order the gravitino mass, $A_\nu \sim \mathcal{O}(m_{3/2})$ \cite{Arkani-Hamed:2000oup}. These interactions enable efficient transfer of the $\phi$ asymmetry to the visible sector before sphaleron freeze-out. 
The resulting baryon asymmetry is given by \Eq{eq:Yphi} with
\bea
R_{\textrm{AD}} = \frac{1}{3}|a_m| \beta \gamma a_{\widetilde{\nu}_R} \sin\theta 
\eea
in terms of $a_{\widetilde{\nu}_R} \simeq m_{3/2}/m_{\widetilde{\nu}_R}$, and the remaining parameters are defined in \Eq{eta4}. 

The dark matter abundance arises from UV freeze-in production of gravitinos, as described in Section~\ref{1234}. In \Fig{fig:Trhm32}, we show that $\Omega_\textrm{DM}/\Omega_\textrm{B}$ is almost independent on $\Trh$ as expected from \Eq{ratio}, up to the running of gauge couplings, and the parameters that give the observed values of $\Omega_\textrm{DM}$ and $\Omega_\textrm{B}$ at given $R_{\rm AD}$ and   $m_{3/2}/m_{1/2}$.

If the RH sneutrino $\widetilde{\nu}_R$ is the NLSP,  the MSSM superpatners will quickly decay to the RH sneutrino via the channel $\tilde\nu_L \to \widetilde{\nu}_R + h$ from \Eq{sneutrinoInt}. 
As a result, BBN constraints are largely evaded because the visible energy $E_{\rm vis}$ is negligible since most of RH sneutrinos decay via  $\tilde\nu_R \to \psi_{3/2}  + \nu_R$ or 
$\tilde\nu_R \to \psi_{3/2} + \nu_L$ arising from mixing between $\tilde\nu_L$ and $\tilde\nu_R$. While highly energetic neutrinos can alter the ratios of light elements, the impact is typically very small. In particular for $\tau_{\tilde\nu_R}< 10^{-3}$ the limits on $m_{\tilde\nu_R}Y_{\tilde\nu_R}$ are very weak (see e.g.~\cite{Kanzaki:2007pd,Hambye:2021moy,Kim:2022gpl}). 

Given that observational limits from decays can typically be satisfied, the most pressing issue is to ensure that NLSP decays are not the dominant production mechanism of gravitinos (which breaks the connection to UV freeze-in). To assess this, we must calculate the freeze-out abundance of the RH sneutrino via \Eq{sneutrinoInt}. Parameterising the  mixing between $\tilde\nu_L$ and $\widetilde{\nu}_R$ following \cite{Belanger:2010cd} we write 
\bea 
\tilde\nu_1 &=&  \tilde\nu_R  \cos \theta_{\tilde\nu} -  \tilde\nu_L \sin \theta_{\tilde\nu},\nonumber\\
\tilde\nu_2 &=& \tilde\nu_R \sin\theta_{\tilde\nu}  + \tilde\nu_L \cos\theta_{\tilde\nu} , 
\eea 
where 
\bea 
\sin\theta_{\tilde\nu}  = \frac{\sqrt{2}A_{\nu} v \sin\beta}{m_{\tilde\nu_2}^2 - m_{\tilde \nu_1}^2}~,
\eea 
with the squared mass matrix
\bea 
M^2 = \left(\begin{array}{cc} m_{\tilde L}^2 + \frac{1}{2} m_Z^2 \cos 2\beta & \frac{1}{\sqrt{2}}A_\nu v \sin\beta \\ 
	\frac{1}{\sqrt{2}} A_\nu v\sin \beta & m_{\tilde\nu_R}^2 \end{array}\right)~.
\eea 
The RH-like sneutrino $\tilde\nu_1$ couplings being those of the LH sneutrino suppressed by a factor $\sin^2\theta_{\tilde\nu}$ due to mixing. We take the limit where $m_{\tilde L} = {\cal O}(5-10) m_{\tilde\nu_R}$ such that $m_{\tilde \nu_1} \sim m_{\tilde\nu_R}$ and $m_{\tilde \nu_2} \sim m_{\tilde L}$. 

There are several channels for annihilation 
$\tilde\nu_1 \tilde \nu_1 \to \nu\nu$ through neutralino t-channel, $\tilde\nu_1 \tilde\nu_1^* \to f \bar f$ via $Z$ exchange, and $ \tilde\nu_1 \tilde\nu_1^* \to b \bar b$ via Higgs exchange. The cross-sections for both channels are proportional to $\sin^4\theta_{\tilde\nu}$.
Because of high power dependence on $\sin\theta_{\tilde\nu}$ it should not be too small for RH sneutrinos to be in thermal equilibrium. 
Depending on parameters, we expect $\sin\theta_{\tilde\nu} \gtrsim {\cal O}(0.01)$. 
For reference, the reader is referred to the figures of \cite{Arina:2007tm}.

The density of $\tilde\nu_1$ prior to decays to the gravitino LSP is set by the freeze-out mechanism. The density of $\tilde\nu_1$ after freeze-out for $m_{\tilde\nu_1} \gtrsim m_h$ is given by \cite{Thomas:2007bu},
\beq
m_{\tilde\nu_1}Y_{\tilde\nu_1} \sim 4 \times 10^{-10} \GeV \left(\frac{0.06}{\sin\theta_{\tilde\nu}}\right)^4 \left(\frac{5m_{\tilde\nu_1}}{m_{\tilde L}}\right)^4.
\eeq
Note that $mY \sim 4 \times 10^{-10} \GeV$ corresponds $\Omega h^2 \sim 0.1$ if it were cosmologically stable.

We require that  $\tilde\nu_1$ decays lead to a negligible increase in the gravitino relic density and that $\tilde\nu_1$ decay does not change BBN. 
Since each sneutrino decay produces a single LSP, the contributions to $\Omega_{\rm DM}$ are diluted by the ratio  of masses $m_{3/2}/m_{\tilde\nu_1}$ according to eq.~(\ref{nlsp}). It follows that for NLSP decays to have a negligible increase in the gravitino relic density, we require 
\beq
m_{3/2} Y_{\tilde\nu_1} \lesssim 10^{-10} \GeV.
\eeq
Furthermore, comparing to the BBN bounds \cite{Kanzaki:2007pd,Hambye:2021moy} these are satisfied for 
\beq
m_{\tilde\nu_1}Y_{\tilde\nu_1} \sin^2\theta_{\tilde\nu} \lesssim 10^{-10} \GeV.
\eeq
 The benchmark model $\sin\theta_{\tilde\nu} = 0.1$ and $m_{\tilde L} = 5 m_{\tilde \nu_1}$ satisfies both requirements.
\Fig{fig:Trhm32} illustrates how $\Omega_\textrm{B}$ and  $\Omega_{\rm DM}$ change in this model as the LSP mass $m_{3/2}$ and $\Trh$ are varied. Observe that the ratio of $\Omega_{\rm DM}/\Omega_\textrm{B}$ is almost constant with $\Trh$, which only occurs since Q-balls and NLSP decays have a negligible impact on $\Omega_\textrm{B}$ and  $\Omega_{\rm DM}$.

\subsection{R-partity violating Affleck-Dine field} \label{sec:RPVAD}

\begin{figure*}[t]
\centering
\includegraphics[height=0.39\textwidth]{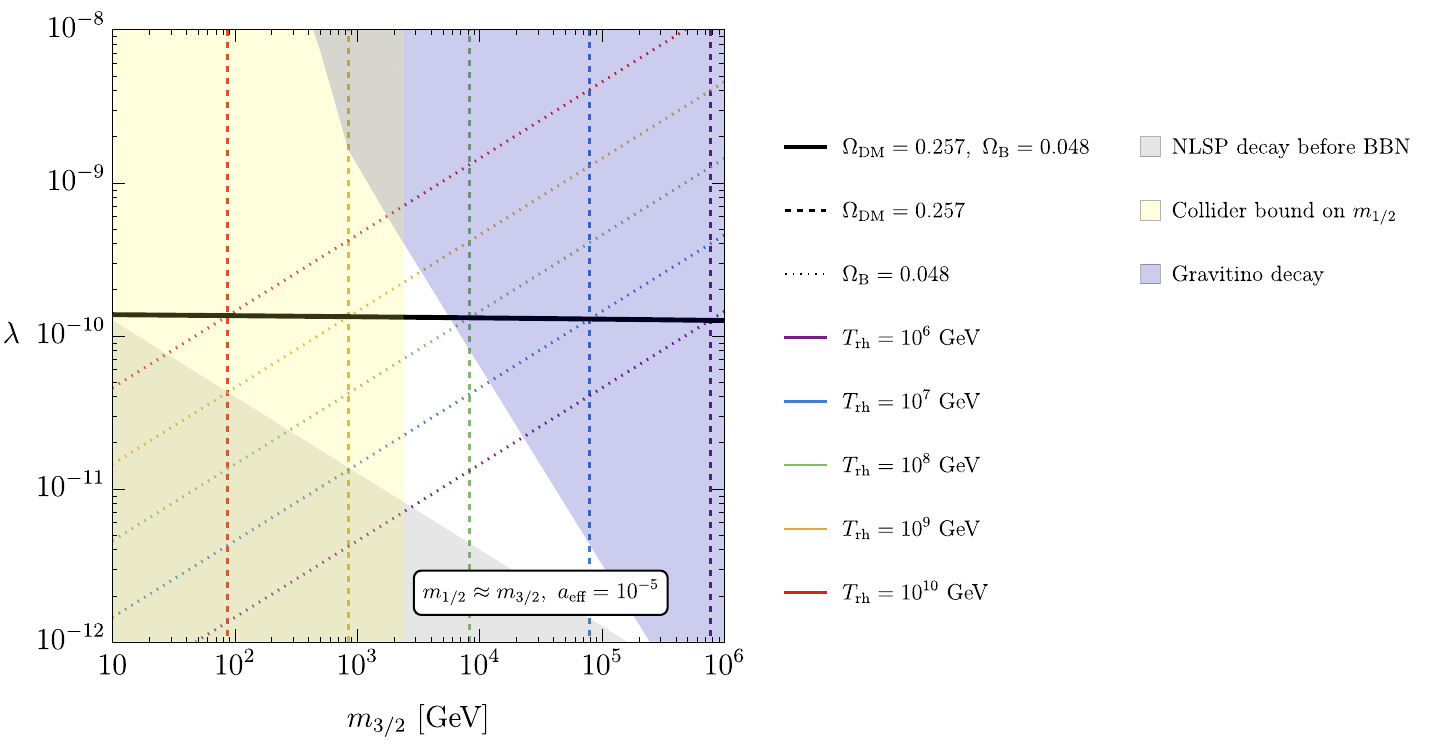}   \hspace{0.35cm}  \\
\includegraphics[height=0.39\textwidth]{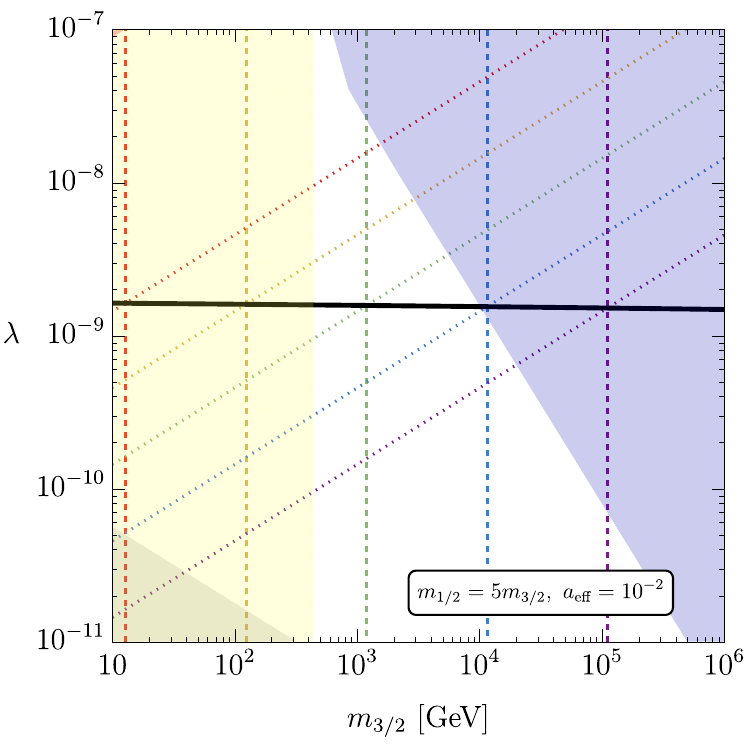}
\includegraphics[height=0.39\textwidth]{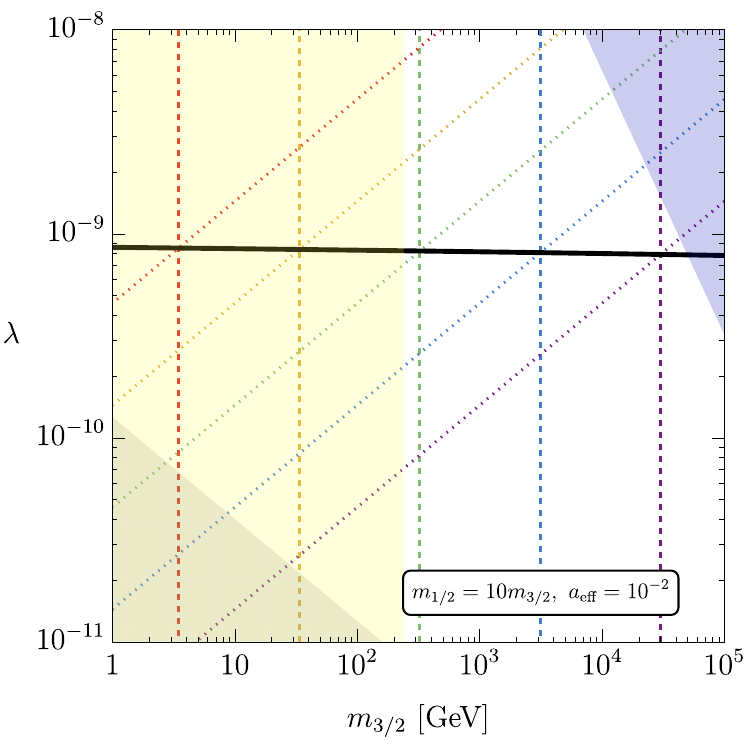}
\caption{We consider the RPV flat direction such that the baryon asymmetry is set by RPV coupling $\lambda$ in eq.~(\ref{RpvAD}), and the parameter $a_{\rm eff}$ implictly defined in \Eq{eq:aeff}. We indicate parameter values that correctly reproduce the observed dark matter abundances (dashed lines) and baryon asymmetry (dotted lines) and via UV freeze-in of the gravitino and Affleck-Dine baryogenesis, respectively. The black solid line identifies both dark matter and baryon abundance are the observed values. Observational constraints are overlaid. The blue shaded region indicates the gravitino LSP lifetime does not lead to indirect detection constraints, and the grey region corresponds to where large $\lambda$ leads to washout of the asymmetry, both taken from \cite{Monteux:2014tia}. While BBN limits can arise from NLSP (with mass $m_{\rm soft}$) decays to gravitinos, for the whole plot area shown, these are found not to be constraining. The yellow region indicates typical LHC limit on coloured superpartners, we take the conservative bound  $m_{1/2}>2.4$ TeV \cite{ATLAS:2024lda}. In the three panels we consider both gravity and gauge mediated type spectra, and the case in which the $\mathcal{O}(1)$ coefficients $R_{\rm AD}$ in the Affleck-Dine model are less than unity. Observe that viable scenarios exist in all cases, although the gravity mediated model with $R_{\rm AD}=1$ is most tightly constrained, requiring $\Trh\sim10^9$ GeV.}
\label{fig:RPVfig}
\end{figure*}

To avoid unacceptable proton decay in the MSSM R-parity is typically imposed, however, R-parity violating (RPV) operators can be introduced if done with care. For the scenario we are considering RPV plays several useful roles: it potentially allows the NLSP to decay prior to BBN, and it reduces the branching fraction to gravitinos such that NLSP decays give only a negligible contribution to the gravitino abundance. 

Moreover, the consideration of R-parity violation  can lead to new opportunities for implementing the Affleck-Dine mechanism. Since the coefficient of these operators can be small and technically natural, these models circumvent many of the issues we have discussed above. 
Higaki et al.~\cite{Higaki:2014eda} studied a scenario involving $\lambda''_{ijk} u^c_i d^c_j d^c_k$ (See also \cite{Harigaya:2014tla}).
Taking the R-parity violating superpotential  
\beq
W = \frac{1}{3!} \lambda_{ijk}'' {\mb U}_i {\mb D}_j {\mb D}_k ~.
\eeq
This can be recast in terms of the Affleck-Dine superfield $\Phi$  as 
\beq \label{RpvAD}
 W = \frac{1}{3} \lambda \Phi^3.
\eeq
It follows that the potential for $\phi$, the scalar component of $\Phi$, is given by\footnote{This can be matched to the general potential of eq.~(\ref{eta}), with the identification $\frac{1}{\Lambda^{n-3}} \to \lambda$. Simply taking  $n = 3$ sets $\lambda=1$.}
\beq
V(\phi) = (m_\phi^2 - c H^2) |\phi|^2 + \frac{\lambda}{3} A \phi^3 + \text{h.c.} + \lambda^2 |\phi|^4.
\eeq
For $V(\phi)$, the cubic potential is a subdominant effect for the expectation value of $\phi$. Then, for $cH^2 \gg m_\phi^2$, the equations of motion for $\phi$ imply
\beq
\phi^* (cH^2 + 2\lambda^2 |\phi|^2) \simeq 0~,
\eeq
and thus
\beq\label{expect}
\langle |\phi|^2 \rangle \sim \frac{c H^2}{2 \lambda^2}~.
\eeq
The potential is effectively flat when $\lambda \ll 1$.   This is technically natural, since R-parity is restored for $\lambda\rightarrow0$. 
 To avoid washing out $B$ asymmetries, one requires $\lambda \lesssim10^{-7}$. Moreover, Q-balls formation is negligible for \cite{Higaki:2014eda}
\begin{equation}
10^{-11} \left(\frac{f_k}{10^{-5}} \right)^2
\lesssim \lambda \lesssim 
 10^{-9} \left( \frac{c_k}{1} \right) \left( \frac{f_k}{10^{-5}} \right)^2 ~.
\end{equation}
where we omit some $\mathcal{O}(1)$ factors and take $\alpha_3\sim0.1$.
This is related to the thermal mass correction is parameterised by $\Delta V \simeq c_k f_k^2 T^2 |\phi|^2$ where $c_k = \mathcal{O}(0.1$--$1)$ is a constant determined by the field degrees of freedom, and $f_k$ is the coupling strength of $\phi$ to the bath states, characteristically $f_k\sim\mathcal{O}(10^{-5}$--$1)$ \cite{Asaka:2000nb,Higaki:2014eda}.

We can represent the scalar potential as radial and angular fields $\phi = \frac{1}{\sqrt{2}} f e^{i\theta}$ in the following manner
\beq
V(f, \theta) = \frac{1}{2} (m_\phi^2 - cH^2) f^2 + \frac{\lambda^2}{4} f^4 + \frac{A}{3 \sqrt{2}} \lambda f^3 \cos \theta~.
\eeq
When $\sqrt{c} H \sim m_\phi$ and $f \sim \frac{m_\phi}{\lambda}$ the angular field $a = f\theta$ has the potential 
\beq
V(a) = \frac{A}{3\sqrt{2}} \lambda f^3 \cos \frac{a}{f} \simeq \frac{A}{3\sqrt{2}} \lambda f^3 \frac{a^2}{f^2} + \dots 
\eeq
We can interpret these quadratic terms as a mass term $\frac{1}{2} m_\theta^2 a^2$ implying 
\beq m_\theta^2 \sim A \lambda f \sim A m_\phi.
\eeq
Therefore, when $c H^2 \sim m_\phi^2$, and when the angular field starts to roll, one has 
\beq
\dot{\theta} \sim \frac{m_\theta^2}{H} \sim \frac{m_\theta^2}{m_\phi} \sim \sqrt{c A}.
\eeq
It follows that the $\phi$ asymmetry at $\Trh$ is given by
\bea  \label{eq:UDDYphi}
Y_\phi 
&= \left( \frac{f^2 \dot{\theta}}{4 H^2 M_{\text{Pl}}} \right) \left( \frac{\Trh}{M_{\text{Pl}}} \right) 
 \sim \sqrt{c^3} \left( \frac{A}{\lambda^2 M_{\text{Pl}}} \right) \left( \frac{\Trh}{M_{\text{Pl}}} \right),
\eea
which gives
\bea\label{eq:aeff}
R_{\rm AD} = a_\textrm{eff} \frac{m_{3/2}}{\lambda^2 \Mpl},
\eea
where $a_\textrm{eff}$ collects the coefficients including $a_m \equiv \frac{A}{m_{3/2}}$.

This $\phi$ asymmetry leads to a baryon asymmetry, and in turn the late time baryon abundance $\Omega_\textrm{B}$.
Accordingly, if we assume the gravitino is the LSP for which the late time abundance is due to freeze-in (as in Section \ref{1234}), then we can compute the ratio of $\Omega_\textrm{DM}$ to  $\Omega_\textrm{B}$ can be obtained from \Eq{ratio}.
For gravity mediation  $A \sim m_{3/2}\sim m_{1/2}$, whereas  for gauge mediation $A \sim m_{1/2}\gg m_{3/2}$, implying two cases
\beq
\frac{\Omega_{\rm DM}}{\Omega_\textrm{B}} \sim \left(\frac{\lambda}{10^{-9}}\right)^2 \times \begin{cases}
~~~1&~~~ {\rm Gravity~med.}\\
\frac{m_{1/2}}{m_{3/2}}  &~~~ {\rm Gauge~med.}
 \end{cases}
 \eeq
where we have used the form of $R_{\textrm{UV},3/2}$ from eq.~(\ref{24}). Thus we conclude that this R-parity violating operator provides an elegant example of our case of interest in which the ratio of dark matter to baryon abundance is $\Trh$ independent, with both Q-balls and washout are avoided due to the (technically natural) small coupling $\lambda\sim10^{-10}-10^{-9}$ as given by Figure~\ref{fig:RPVfig}.

Note that an expectation value for ${\mb U}_i {\mb D}_j {\mb D}_k$ breaks U(1)$_Y$ and SU(3)$_c$.  A large $\langle \phi \rangle=f(t)$ implies large masses for the U(1)$_Y$ and SU(3)$_c$ gauge bosons which can impede thermalization of the visible sector~\cite{Allahverdi:2005mz}. 
Therefore, the basic condition for thermalization to proceed as usual is that the gluon and gluino masses, $M_g \sim g_3 f(t)$, become smaller than the temperature at $T = T_{\rm rh}$. If this condition is satisfied, the effect of the gauge boson mass becomes negligible during the radiation-dominated era, since $f(t) \propto a^{-3/2}$. 

Let us briefly examine this condition, focusing on its parametric dependence. At $T = T_{\rm rh}$, the ratio of the energy density of the Affleck-Dine (AD) fields to the total energy density is given by
\begin{equation}
\frac{\rho_\phi}{\rho_{\rm tot}} \sim \frac{f^2(t_{\rm AD})}{M_{\rm Pl}^2} \sim \frac{m_\phi^2 f^2(t_{\rm rh})}{T_{\rm rh}^4}.
\end{equation}
From this expression, we can derive the ratio between $f(t_{\rm rh})$ and $T_{\rm rh}$ as
\begin{equation}
\frac{f(t_{\rm rh})}{T_{\rm rh}} \sim \frac{f(t_{\rm AD}) T_{\rm rh}}{m_\phi M_{\rm Pl}} \sim \frac{T_{\rm rh}}{\lambda M_{\rm Pl}}.
\end{equation}
Here, $\lambda$ and $T_{\rm rh}$ are constrained by the baryon-to-dark matter ratio as well as by their absolute values. For a given value of $\lambda$, the reheating temperature $T_{\rm rh}$ can be fixed by the model parameters and cosmological abundances using eq.~(\ref{eq:UDDYphi}). Consequently, the ratio between the gauge boson mass and the temperature at $T = T_{\rm rh}$ is expressed as
\begin{equation}
\left.\frac{M_g}{T}\right|_{T = T_{\rm rh}} \sim \left( \frac{Y_\phi}{10^{-10}} \right) \left( \frac{\lambda}{10^{-9}} \right) \left( \frac{0.1\,{\rm GeV}}{A} \right).
\end{equation}
One may be concerned that for $M_g/T\gg1$ that the connection between the dark matter and baryons will be disrupted, thus we restrict to $M_g/T\ll1$.
In the parameter space considered in our model, this ratio can be sufficiently small. Thus, the gravitino DM production proceeds via the usual UV freeze-in mechanism.

Various concerns arise once the Lagrangian includes RPV operators, and we discuss them in turn below. Since we have seen in the above that gauge mediated models with a stau NLSP can decay without dominating the gravitino abundance or disrupting BBN, we will restrict our discussion in this section to gravity mediated scenarios with $m_{\rm NLSP}\sim m_{3/2}$.
Suppose that the neutralino is the NLSP, its lifetime due to the presence of the $UDD$ operator has \cite{Higaki:2014eda,Dreiner:2025kfd}
\beq
\tau_{{\rm NLSP}}^{\not R_p}
\simeq 0.1\sec
\left(\frac{10^{-10}}{\lambda }\right)^2
\left(\frac{1 {\rm TeV}}{m_{\rm NLSP}}\right)^5\left(\frac{2 {\rm TeV}}{m_{\tilde q}}\right)^4,
\eeq
where $m_{\tilde q}$ is the common scale of the squarks.
Note that if the RPV coupling is too large it can lead to wash-out of generated baryon asymmetry, following  \cite{Dreiner:1992vm,Buchmuller:2007ui}, to avoid this we restrict  $\lambda \lesssim10^{-7}$. Observe that for the parameter values shown the NLSP decays to Standard Model particles via the RPV operator well before BBN.

RPV also implies that the gravitino is now unstable and we require that its lifetime is sufficiently long that its decays are not observable. Since the decay rate is suppressed by both the Planck scale and the RPV coupling (which is assumed to be small) the gravitino typically always lives longer than the age of the Universe. Specifically, consider we only introduce the RPV operator of $UDD$, the lifetime of the  gravitino LSP (with stau NLSP) can be estimated to be \cite{Moreau:2001sr,Buchmuller:2007ui,Monteux:2014tia,Monteux:2014hua}
\bea 
\tau_{\psi_{3/2}}
&\simeq 4.6\times 10^{29}\sec
\left(\frac{10^{-10}}{\lambda }\right)^2
\left(\frac{1 {\rm TeV}}{m_{3/2}}\right)^7
\left(\frac{m_{\tilde q}}{2 {\rm TeV}}\right)^4.
\eea
To satisfy constraints from indirection searches restricts $\tau_{3/2} \gtrsim  10^{27}\sec$  \cite{Albert:2014hwa,Ando:2015qda,HAWC:2017udy}, which is not overly constraining.  
All those constraints are shown in Figure~\ref{fig:RPVfig}.

We close by noting that care should be taken not to introduce both $B$ and $L$ violating dimension four or five operators, as this will generally lead to unacceptably fast proton decay. Moreover, if the gravitino is lighter than the proton, then $B$-violation alone can lead to proton decay which can run into tension with constraints \cite{Choi:1996nk}.

\section{Conclusions}
\label{Sec7}
This work has highlighted that the Affleck-Dine mechanism is highly symbiotic with UV freeze-in production of dark matter. Through pairing these mechanisms, the abundances of both the baryons and dark matter can be simultaneously explained through an appropriate value of the reheating temperature of the Universe $\Trh$. In particular, for UV freeze-in via a dimension five operator, then the ratio $\Omega_{\rm DM}/\Omega_\textrm{B}$ does not exhibit any dependency on the reheating temperature of the Universe.  While this is more generally true for any mechanism in which an observed quantity has a linear dependence of $\Trh$, these two mechanisms provide the cleanest examples.

Since the Affleck-Dine mechanism is intrinsically supersymmetric (to obtain flat directions), we identified a number of SUSY candidates that are ideal for realizing this scenario.
In particular, we have seen that the gravitino provides an excellent candidate for linking UV freeze-in and Affleck-Dine baryogenesis. Figs.~\ref{fig:TrhOmega} \& \ref{fig:Trhm32} nicely illustrate the general scaling principles that lead to the parametric coincidence between $\Omega_{\rm DM}$ and $\Omega_\textrm{B}$.

A number of constraints arise from consistency conditions and observational limits.  In particular, to maintain the connection between $\Omega_{\rm DM}$ and $\Omega_\textrm{B}$, one must avoid altering either quantity through other processes, in particular, Q-ball formation and NLSP decays. There may also be BBN constraints from NLSP decays. We have outlined a range of scenarios in which the various constraints can be satisfied, including specific models in which many issues are avoided. In particular, we have highlighted the case in which RH sneutrino is identified as the NLSP and the Affleck-Dine field as particularly attractive. 

It is notable that by requiring the reheat temperature to be fundamental in establishing both the baryon asymmetry and the dark matter abundance, it leads to a more constrained scenario than either of these two UV-dominated mechanisms seperately.

\vspace{5mm}
{\bf Acknowledgements.}~This manuscript has been authored in part by Fermi Forward Discovery Group, LLC under Contract No.89243024CSC000002 with the U.S. Department of Energy, Office of Science, Office of High Energy Physics. CSS is supported by  the National Research Foundation of Korea (NRF) grant funded by the Korea government (MSIT) RS-2022-NR072128 and IBS-R018-D1. JU is supported by NSF grant PHY-2209998.

\appendix

\section{Avoiding Matter Domination}
\label{ApA}

To maintain our link between UV freeze-in and baryogenesis it is important to ensure that the energy density stored in the flat direction $\phi$ does not come to dominate the energy density of the Universe prior to its decay. In this appendix, we derive a simple constraint on the initial field value $\langle \phi \rangle$ to avoid this issue.
We assume that the flat direction begins oscillating when the Hubble parameter drops to $H \sim m_\phi$, at which point its energy density is
\begin{equation}
\rho_\phi \sim m_\phi^2 |\phi|^2~.
\end{equation}
At the same time, the background inflaton energy density is approximately $\rho_I \sim M_{\rm Pl}^2 H^2 \sim  M_{\rm Pl}^2 m_\phi^2$. Hence, the initial ratio is (cf.~(eq.~(\ref{eq10}))
\begin{equation}
\frac{\rho_\phi}{\rho_I} \sim \frac{f(t_{\rm AD})^2}{M_{\rm Pl}^2}~.
\end{equation}
This ratio maintains until reheating. Since the Affleck-Dine field redshifts as matter $\rho_\phi \propto a^{-3}$, while for the radiation $\rho_r \propto a^{-4}$ after reheating, the ratio at $T=T_*$, the temperature $\phi$ condensate dissapates (by thermalisation or $\phi$ deacy), is
\begin{equation}
\left.\frac{\rho_\phi}{\rho_r}\right|_{T=T_*} \sim~ \frac{f(t_{\rm AD})^2}{M_{\rm Pl}^2} \frac{\Trh}{T_*}~, 
\end{equation}
where $T_*$ denotes the temperature where $\phi$ decays or scatters with the Standard Model bath, whichever is larger. 
$T_*$ can be obtained from $\Gamma_\phi \sim H$, where $\Gamma_\phi$ is the decay or scattering rate of $\phi$. 

We can check this criterion in the specific context of the models of Section \ref{Sec6}.
We find for the sneutrino model, $\Gamma_\phi \sim A_\nu^2/T$, which gives $T_* \sim (\Mpl A_\nu^2)^{1/3}$. To avoid early matter domination, we require
\begin{equation}
A_\nu \gtrsim \frac{f(t_{\rm AD})^3 \Trh^{3/2}}{\Mpl^{7/2}}~,
\end{equation}
which is satisfied for the parameter space we consider. Similarly, $\Gamma \sim \alpha_s^2 T$ for the RPV model. We get $T_* \sim \alpha_s^2 \Mpl$ so $\phi$ thermalizes quickly with the Standard Model bath. Hence in either case matter domination does not occur for the parameter space of interest.

\section{Production before reheating}
\label{ApB}
In this appendix, we discuss a subtlety regarding non-instantaneous reheating. Dark matter production may initiate once the thermal bath is populated and while the inflaton (or other field) is still injecting energy into the bath. Indeed, in certain cases, the thermal bath may reach temperatures $T_{\rm max}$ higher than $\Trh$ before subsequently cooling \cite{Chung:1998rq,Giudice:2000ex}. Thus we should ascertain whether dark matter production at temperatures $T>\Trh$ (times $t<\Trh$) can be typically ignored.

Consider the Boltzmann equation $\dot{n} + 3Hn = \Gamma(T)$ where we parameterize the rate $\Gamma(T) \sim R T^6/\Lambda^2$. For a given Hubble time $\Delta t = H^{-1}$ with $t < \Trh$ the change in the number density $\Delta n$ is parametrically
\beq
\Delta n \sim \Gamma(T) \frac{1}{H}.
\eeq
We can estimate the yield prior to freeze-in at $t = \Trh$
\beq
\label{666}
\left(\frac{\Delta n}{s}\right)_{\Trh}  = \Trh \left(\frac{\Delta n}{\rho_{\rm tot}}\right)_{t = \Trh}.
\eeq
Because  for $t < \Trh$ one has $\Delta n \propto a^{-3}$, and $\rho_{\rm tot}  \propto a^{-3}$, it follows 
\beq
\left(\frac{\Delta n}{\rho_{\rm tot}}\right)_{t = \Trh} \simeq \left(\frac{\Delta n}{H^2 M_{\rm Pl}^2}\right)_{t < \Trh}
\simeq \left(\frac{\Gamma(T)}{H^3 M_{\rm Pl}^2}\right).
\eeq
For $t < \Trh$ one has $\rho_{\text{rad}} \ll \rho_{\rm tot} \sim H^2 M_{\rm Pl}^2$ and thus $T^4 \ll H^2 M_{\rm Pl}^2$.
Therefore for $T>\Trh$ the Hubble rate is
\beq
\label{eqeq}
H(T)\simeq  \frac{T^2}{M_{\rm Pl}} \times \left(\frac{T}{\Trh}\right)^{m} ~~~~~(m> 0).
\eeq
where $m$  parameterizes the expansion rate prior to reheating (for standard inflationary reheating, one typically takes $m=2$  cf.~\cite{Giudice:2000ex}). Then
using eq.~(\ref{eqeq}) in eq.~(\ref{666}) one obtains
\bea
\frac{\Delta n}{s}\bigg|_{\Trh} &= \Trh \left(\frac{\Gamma(T)}{H^3 M_{\rm Pl}^2}\right) \\
&\sim \Trh \left(\frac{\Gamma(T) M_{\rm Pl}}{T^6}\right) \times \left(\frac{\Trh}{T}\right)^{3m}.
\eea
Alternatively, we can write $(T/\Trh)^{m}\simeq \sqrt{\rho_{\rm tot}/\rho_{\rm rad}}$.
Taking the production rate to be  $\Gamma(T) \sim R \frac{T^6}{\Lambda^2}$ one has
\beq
\frac{\Delta n}{s} = R \frac{\Trh M_{\rm Pl}}{\Lambda^2} \times \left(\frac{\rho_{\rm rad}}{\rho_{\rm tot}}\right)^{3/2}.
\eeq
While there is a $\Trh$-dependence, for temperatures in excess of the reheat temperature at times $t<\Trh$, this leads to a suppression since $\rho_{\rm rad}\ll \rho_{\rm tot}$. Thus dark matter production during $t<\Trh$  can typically be neglected.

\section{More General NLSP Decays}
\label{NLSP-Decays}

Considering the case of a gravitino LSP whose 
relic abundance is given by eq.~({\ref{o32}) in this appendix
we examine the implications of NLSP decays for typical candidates. These decays may either invalidate eq.~({\ref{o32}) or lead to experimental exclusions. Since there are a number of differences between gravity mediation and gauge mediation, such as the sparticle spectra and the conditions for avoiding Q-ball formation, we must consider these two scenarios separately.

\subsection*{B1:~Gauge mediated SUSY breaking}
\label{3D}

In the case of gauged mediated supersymmetry breaking the gravitino is generically the LSP and can be made arbitrarily light compared to the mass scale of the other superpartners $m_{\rm soft}$ (up to cosmological consideration).\footnote{A similar mass hierarchy  $m_{3/2}\ll m_{\rm NLSP}$ can arise in a special class of  UV completions to supersymmetry called ``no-scale`` supergravity  \cite{Cremmer:1983bf,Ellis:1984kd}, which involves a non-minimal K\"ahler potentials. Leading to similar conclusions regarding $\Delta\Omega_{3/2} h^2$.}   
The NLSP is not uniquely identified, although it is often taken to be the neutralino. Identifying the NLSP is especially important  in this case since it will decay to the gravitino LSP and its contribution to the gravitino abundance is given by
\beq 
\Delta\Omega_{3/2} h^2 = \frac{m_{3/2}}{m_{\text{NLSP}}} \Omega_{\text{NLSP}} h^2~.
\eeq
Thus unless the NLSP is especially poor at annihilating, then on anticipates that $\Delta\Omega_{3/2} h^2\ll1$ for a modest splitting $m_{3/2}\ll m_{\text{NLSP}}$. Such a mass hierarchy between the gravitino and other sparticles is expected in gauge mediation, cf.~eq.~(\ref{soft}). 

To verify this assertion that one can typically neglect contributions to $\Omega_{3/2} h^2$ provided $m_{3/2}\ll m_{\text{NLSP}}$, let us consider a couple of examples. Suppose the NLSP is a neutralino $\chi$, such a candidate is a natural WIMP candidate over much of its parameter space as such $\Omega_{\text{NLSP}} h^2\sim 0.25$ for $m_{\chi}\sim1$ TeV. Then the requirement that $\Delta\Omega_{3/2} h^2 \ll1$ implies $m_{3/2}\ll 1$ TeV or equivalently, $M\ll M_{\rm Pl}$ which is a typical expectation.
Let us consider  the stau NLSP as another example
 \cite{Okada:2007na}
\beq \label{stau}
\Delta\Omega_{3/2} h^2 = 0.01 \left(\frac{m_{3/2}}{1   \text{GeV}}\right)  \left(\frac{m_{\tilde{\tau}}}{1   \text{TeV}}\right)~.
\eeq
The parameter values indicated above would imply a percent level change in the relic gravitino abundance.
Thus this hierarchical separation between $m_{3/2}$ and $m_{\rm soft}$ allows the contribution to the relic gravitino abundance from NLSP decays to be negligible in many cases.

\subsection*{B2:~Gravity Mediated SUSY Breaking}

We next consider the scenario in which supersymmetry breaking is communicated to the visible sector only via gravity. 
We will restrict our attention to leptonic flat directions (e.g.~$\phi/\sqrt{2}= H_u=L $ or $\phi=\widetilde{\nu}_R$) in order to avoid Q-ball formation as discussed in Section \ref{2B}.
In this case the gravitino can be arranged to be the LSP, as discussed in Section \ref{Sec3}, and then identifying the NLSP is important for understanding the phenomenology and constraints.

In gravity mediated models the low energy spectrum of superpartners is determined by UV boundary conditions and renormalisation group running. The archetypal model of gravity-mediated supersymmetry is the Constrained Minimal Supersymmetric Standard Model (CMSSM), which enforces a number of conditions on the soft parameters at the GUT scale. In particular, one assumes universal soft SUSY-breaking masses for the scalars $m_0$, gaugino $m_{1/2}$, and universal trilinear supersymmetry-breaking A-terms ($A_0$).  
After running to low energy, the lightest Standard Model superpartner is found to be either the (mostly bino) neutralino $\chi$ or the stau $\widetilde{\tau}$, with $m_\chi\simeq0.4m_{1/2}$ and $m_{\widetilde{\tau}}^2\simeq m_0^2+0.15m_{1/2}^2$ \cite{Martin:1993ft}.  One of these two states will typically be the NLSP (if the gravitino is the LSP); however, as we discuss below, there are caveats.

The problem arises that neither the bino or the stau annihilate efficiently enough to heavily deplete their relic abundances. Moreover, without significant tuning, the gravitino cannot be made significantly lighter than the NLSP in gravity mediation \cite{Kersten:2009qk}: $m_{\rm NLSP}\sim m_{3/2}$. By inspection of eq.~(\ref{stau}) for the stau one finds that the contribution to the gravitino abundance from NLSP decays is typically comparable to (or greater than) the thermal gravitino abundance, i.e.~$\Delta\Omega_{3/2}h^2\not\ll \Omega_{3/2}h^2$. A similar conclusion can be found for the bino-like neutralino \cite{Ellis:2003dn,Roszkowski:2004jd}. This is clearly a concern since we are relying on $\Trh$-dependent thermal production of gravitinos to provide the link between $\Omega_{\rm DM}$ and the baryon asymmetry in eq.~(\ref{ratio}), however, this is an issue that can be resolved with some further model building.


The CMSSM is one of the most `constrained' supersymmetric extensions of the Standard Model one can construct (hence its name). Deviations from the universal boundary conditions can change the NLSP. For instance, if one considers non-universal soft
breaking masses for the Higgses at the GUT scale, then depending on the choice of parameters, the NLSP can be neutralino, Higgsino, or sneutrino in addition to a pure Bino or stau \cite{Ellis:2002wv}.

Furthermore, the inclusion of additional states or symmetries can further perturb the spectrum while remaining within the framework of gravity-mediated SUSY breaking.  For instance, one can forbid the $\mu H_uH_d$ operator at tree level via a symmetry and subsequently reintroduce this term via the Giudice-Masiero mechanism. This both solves the $\mu$ problem of the MSSM and allows $\mu$ some degree of freedom \cite{Evans:2014pxa}. If $m_{3/2}<\mu<m_{\rm soft}$, then the Higgsinos can be the NLSP. 

While it remains challenging to obtain $\Delta\Omega_{3/2} h^2\ll0.1$ with $m_{3/2}\not\ll m_{\rm soft}$ and $m_{\rm soft}\sim$ TeV (in order to avoid collider constraints) it is possible in some parts of parameter space. For instance, for a Higgsino NLSP the freeze-out abundance  (prior to decays) is given by \cite{Arkani-Hamed:2006wnf}
\beq
\Omega_{\widetilde{H}} h^2\sim0.1\left(\frac{\mu}{1~{\rm TeV}}\right)^2~.
\eeq
By comparison to eq.~(\ref{nlsp}) this implies
\beq
\Delta\Omega_{3/2} h^2 = 0.01\left(\frac{m_{3/2}}{300~{\rm GeV}}\right)\left(\frac{\mu}{350~{\rm GeV}}\right)~.
\eeq
With the stated parameter values  NLSP decays produce 10\% of the gravitino dark matter, which is acceptable but less compelling. Moreover, one anticipates that the other superpartners (including the coloured states) appear within an order of magnitude of $m_{3/2}$. Thus one predicts sub-3 TeV gluino which is near the edge of experimental tension. This likely also requires some tuning in the supergravity model to get an order-of-magnitude mass separation between the gluino and gravitino \cite{Kersten:2009qk}. 

We have identified three scenarios in which NLSP decays can be neglected: 
\begin{itemize}
\item Resonant NLSP annihilation.
\item Mass splittings $m_{3/2}\ll m_{\rm NLSP}$.
\item R-parity violating decays of the NLSPs.
\end{itemize}
The case of  R-parity violating decays was already discussed in Section \ref{Sec6}.

Many of the  NLSPs discussed above can annihilate to Standard Model states on resonance via the heavy scalar components of the Higgs sector, the  heavy CP-even state $H^0$ and the CP-odd $A^0$. Resonant annihilation can be orders of magnitude more effective at reducing cosmological abundances. 
Cosmological abundances following resonant annihilation have previously been studied in the literature for neutralinos (including Bino-Higgsino) \cite{Beneke:2022rjv,Choi:2006hi}, sneutrinos \cite{Arina:2007tm,Lopez-Fogliani:2021qpq}, and staus \cite{Pradler:2008qc,Heisig:2013rya}. Resonant annihilation can lead to $\Omega_{\text{NLSP}} h^2\ll0.1$ and thus  one can obtain $\Delta\Omega_{3/2}\ll \Omega_{3/2}$ with $m_{3/2}\sim m_{\rm NLSP}$, removing the constraints and disruption of NLSP at the cost of fine-tuning the masses onto resonance: $m_H\approx 2m_{\rm NLSP}$.


\end{document}